\documentclass[prl,superscriptaddress,amsmath,amssymb,twocolumn,nofootinbib]{revtex4-2}

\usepackage{graphicx,bm,amsmath}
\usepackage[usenames,dvipsnames]{xcolor}
\usepackage[colorlinks,bookmarks=false,citecolor=NavyBlue,linkcolor=Red,urlcolor=blue,%
]{hyperref}
\usepackage[percent]{overpic}

\usepackage[bbgreekl]{mathbbol}
\usepackage{xcolor}
\usepackage[normalem]{ulem}
\usepackage{braket}
\usepackage[ruled,vlined]{algorithm2e}
\usepackage{comment}
\usepackage{amsthm}
\usepackage{enumerate}
\usepackage{enumitem}
\usepackage{subfigure}

\def\doi{http://dx.doi.org/}

\newcommand{\be}{\begin{equation}}
\newcommand{\ee}{\end{equation}}
\newcommand{\bec}{\begin{equation*}}
\newcommand{\eec}{\end{equation*}}
\newcommand{\bea}{\begin{eqnarray}}
\newcommand{\eea}{\end{eqnarray}}

\newcommand{\newsec}[1]{\textcolor{blue}{{\textit{#1}.-}}}

\graphicspath{{./figs/}}

\newcommand{\Z}{\mathbb{Z}}
\newcommand{\HH}{\mathcal{H}}
\newcommand{\XX}{\mathcal{X}}
\newcommand{\ZZ}{\mathcal{Z}}

\newcommand{\titleinfo}{Bell sampling in Quantum Monte Carlo simulations}

\newcommand{\Tr}{\text{Tr}}   %

\begin{document}
\title{\titleinfo}

\author{Poetri Sonya Tarabunga}
\email{poetri.tarabunga@tum.de}
\thanks{These authors contributed equally.}
\affiliation{Technical University of Munich, TUM School of Natural Sciences, Physics Department, 85748 Garching, Germany}
\affiliation{Munich Center for Quantum Science and Technology (MCQST), Schellingstr. 4, 80799 M{\"u}nchen, Germany}

\author{Yi-Ming Ding}
\email{dingyiming@westlake.edu.cn}
\thanks{These authors contributed equally.}
\affiliation{State Key Laboratory of Surface Physics and Department of Physics, Fudan University, Shanghai 200438, China}
\affiliation{Department of Physics, School of Science, Westlake University, Hangzhou 310030,  China}
\affiliation{Institute of Natural Sciences, Westlake Institute for Advanced Study, Hangzhou 310024, China}
    
\begin{abstract}
Quantum Monte Carlo (QMC) methods are essential for the numerical study of large-scale quantum many-body systems, yet their utility has been significantly hampered by the difficulty in computing key quantities such as off-diagonal operators and entanglement. This work introduces Bell-QMC, a novel QMC framework leveraging Bell sampling, a two-copy measurement protocol in the transversal Bell basis. We demonstrate that Bell-QMC enables an efficient and unbiased estimation of both challenging classes of observables, offering a significant advantage over previous QMC approaches. Notably, the entanglement across all system partitions can be computed in a single Bell-QMC simulation. We implement this method within the stochastic series expansion (SSE), where we design an efficient
update scheme for sampling the configurations in the Bell basis. We demonstrate our algorithm in the one-dimensional transverse-field Ising model and the two-dimensional $\Z_2$ lattice gauge theory, extracting universal quantum features using only simple diagonal measurements. This work establishes Bell-QMC as a powerful framework that significantly expands the accessible quantum properties in QMC simulations, providing a substantial advantage over conventional QMC. %

\end{abstract}

\maketitle

\newsec{Introduction}
Strongly correlated quantum many-body systems lie at the heart of condensed matter physics, displaying a rich variety of novel phenomena, and holding immense potential for technological breakthroughs. %
However, their inherent complexity presents a formidable computational challenge. %
A wide range of numerical methods have been developed to overcome this hurdle, such as the density matrix renormalization group (DMRG)~\cite{Schollwock2011dmrg,orus2014practical} and quantum Monte Carlo (QMC) techniques~\cite{sandvik2010computational,Gubernatis2016qmcbook,Becca2017}. DMRG has found immense success in simulating one-dimensional systems, yet it struggles to simulate higher-dimensional systems. QMC techniques are not hampered by the dimensionality, making them indispensable tools for accurately studying higher-dimensional quantum many-body systems. While QMC techniques have achieved maturity after decades of development, there are still significant challenges in extracting certain key quantities. In particular, two outstanding challenges are the computation of (1) general off-diagonal observables and (2) nonlinear observables such as entanglement measures, and intense research efforts have been devoted to tackle these challenges.

As a central physical quantity for characterizing quantum many-body systems, entanglement usually follows the area law for gapped ground states~\cite{wolf2008area}, with violations signaling universal information~\cite{calabrese2004entanglement,metlitski2015entanglement,kitaev2006,levin2006,amico2008,Laflorencie2016entReview}.
Existing QMC approaches typically express the R\'enyi-2 entanglement entropy (EE) as a ratio between two replica partition functions $\Tilde{Z}_2/\Tilde{Z}^2$~\cite{calabrese2004entanglement}, which generally cannot be computed through simple measurements of the partition function. 
Even in cases where it is possible, it is difficult to achieve reasonably small statistical errors even in one-dimensional systems~\cite{hastings2010measuring,humeniuk2012quantum}, where the entanglement does not grow significantly. 
This necessitates the implementation of more intricate tricks to reduce the errors, typically adapted from classical Monte Carlo or molecular dynamics, such as the increment trick~\cite{hastings2010measuring,humeniuk2012quantum,ZhouX2024incremental}, thermodynamic integration~\cite{melko2010finitesizescaling}, reweight-annealing algorithm~\cite{ding2024reweight,Wang2024ent}, or non-equilibrium Jarzynski's equation~\cite{emidio2020entanglement}, often at the cost of introducing systematic errors. This type of technique has recently been applied to access off-diagonal observables~\cite{zhiyan2024offdiag}, though calculating certain Pauli operators still suffers from the sign problem.

In this work, borrowing insights from quantum information theory, we introduce a novel QMC framework leveraging Bell sampling, a protocol involving the measurement of two copies of the quantum state in the transversal Bell basis. Bell sampling has been recognized as a powerful tool for efficiently performing various quantum information tasks~\cite{hangleiter2024bell, huang2022quantum,huang2021information,haug2023scalable,haug2024efficient,montanaro2017learning,gross2021schur,grewal2024efficient,king2024triplyefficient}, offering substantial advantage over conventional experiments where each copy of the state is measured separately~\cite{huang2022quantum,huang2021information,chen2022exponential}. It has also been demonstrated experimentally~\cite{bluvstein2024logical, huang2022quantum,haug2023scalable,haug2024efficient,haug2025efficientwitnessingtestingmagic,islam2015measuring,Kaufman2016,Linke2018}. Crucially, it enables the efficient estimation of the set of all Pauli operators~\cite{huang2021information,king2024triplyefficient} and other physical quantities, including entanglement~\cite{horodecki2002method,ekert2002direct}, nonstabilizerness~\cite{gross2021schur,haug2024efficient,haug2023scalable}, and inverse participation ratio~\cite{haug2024pseudorandomunitariesrealsparse}. In contrast, these quantities typically require the number of resources exponential in the system size to estimate with single-copy measurements. Thus, Bell measurements provide exponential advantage over conventional single-copy measurements.

We show that Bell sampling can be integrated into QMC, that we call Bell-QMC, which directly addresses the aforementioned limitations of conventional QMC in a very simple way: all Pauli operators and the R\'enyi-2 EE of any subsystem can be estimated in a single simulation in an unbiased manner. It leads to a remarkably powerful and yet conceptually simple QMC framework, offering a significant advantage over the sophisticated techniques developed in the past decade. Having no counterpart in classical Monte Carlo techniques, Bell-QMC represents a fundamental paradigm shift in the extraction of quantum properties via QMC simulations, directly exploiting the exponential advantage that multi-copy measurements offer over single-copy measurements for accessing certain quantum observables~\cite{huang2022quantum,huang2021information,chen2022exponential}. As a result, Bell-QMC directly provides a significant advantage over conventional QMC.

Focusing on stochastic series expansion (SSE), we design an efficient
update scheme for sampling the configurations in the Bell basis, based on a modification of the standard cluster update algorithm~\cite{sandvik2003sse_ising}. %
We benchmark our method through various examples, including the 1D transverse field Ising model (TFIM) and the 2D $\Z_2$ lattice gauge theory (LGT), demonstrating the power of our algorithm in enabling the direct, accurate, and unbiased measurement of entanglement in large-scale lattices, a capability that was previously unattainable.

\newsec{Brief review of Bell sampling} \label{sec:bell_sampling}
We consider a system of $N$ qubits associated with the Hilbert space $\mathcal{H}_N=\bigotimes_{j=1}^N\mathcal{H}_j$, where $\mathcal{H}_j\simeq \mathbb{C}^2$. We define the Pauli matrices $\sigma_{0 0} = I$, $\sigma_{01} = X$, $\sigma_{10} = Z$, and $\sigma_{11} = iY$. %
The Bell states are defined as $\ket{\sigma_r} = (\sigma_r \otimes I ) \ket{\Phi^+}$ for $r=(r_z,r_x) \in \{0,1\}^2$, where $\ket{\Phi^+} = (\ket{00} + \ket{11})/\sqrt{2}$. Specifically, $\ket{\sigma_{00}} = (\ket{00} + \ket{11})/\sqrt{2}, 
    \ket{\sigma_{01}} = (\ket{10} + \ket{01})/\sqrt{2}, \ket{\sigma_{10}} = (\ket{00} - \ket{11})/\sqrt{2},$ and $\ket{\sigma_{11}} = (\ket{01} - \ket{10})/\sqrt{2}$, which form an orthonormal basis for two-qubit systems, known as the Bell basis. For ease of notation, hereafter we will write $\ket{\sigma_{r^z r^x}}= \ket{r^z,r^x}$.

For two states $\rho_A, \rho_B$, measurements in the Bell basis are performed by preparing the tensor product state $\rho_A \otimes \rho_B$, and then applying the Bell transformation $U_\mathrm{Bell} = \bigotimes_{i=1}^N (\mathrm{H}\otimes {I}_2) \mathrm{CNOT} $ with the Hadamard gate $\mathrm{H}$ and the CNOT gate. We can then measure in the computational basis, yielding a measurement outcome $\mathbf{r}=\{\mathbf{r}^z, \mathbf{r}^x\}$ with $\mathbf{r}^z, \mathbf{r}^x \in \{0,1 \}^{N}$ with probability $P(\mathbf{r}) = \frac{1}{2^N} \Tr[\rho_A \sigma_\mathbf{r} \rho_B^* \sigma_\mathbf{r} ]$. A more detailed review can be found in the Supplemental Material~\cite{supmat}

\newsec{Algorithm: Bell-QMC} \label{sec:algo}
In the standard QMC framework, the partition function $\tilde{Z}=\Tr(e^{-\beta H})$ is sampled in a local basis, which is typically composed by eigenstates of Pauli-$Z$ operators. This can be understood as sampling from single copies of the state, where each basis state $\ket{a}$ is sampled with probability $c_a \propto \bra{a} e^{-\beta H} \ket{a}$. The main idea of Bell-QMC is to shift from single-copy measurements to two-copy measurements in the Bell basis.

The algorithm formulates SSE in the Bell basis, thus effectively performing Bell sampling. The partition function of two copies of the thermal state is $\Tilde{Z}_B =\Tr\left[e^{-\beta H} \otimes e^{-\beta H} \right]  = \Tr\left[e^{-\beta \HH}\right]$,
where $\HH = (H \otimes I + I \otimes H)$ is a two-copy Hamiltonian. 
For concreteness, we now consider the TFIM with 
\begin{equation} \label{eq:ising}
    H = -\sum_{\langle i,j \rangle} Z_i Z_j - h\sum_i X_i,
\end{equation}
and the corresponding two-copy Hamiltonian is $\HH = - 2\sum_{\langle i,j \rangle} \ZZ_{\langle i,j \rangle}  - 2h\sum_i \XX_i$, where $\XX_i = \frac{1}{2}(X_i \otimes I_i + I_i \otimes X_i)$ and $\ZZ_{\langle i,j \rangle} = \frac{1}{2}(Z_i Z_j \otimes I_i I_j + I_i I_j \otimes Z_i Z_j)$. The action of these operators on a Bell state is given by 
\begin{align} 
    &\XX_i \ket{r^z_i,r^x_i} = \delta_{r^z_i,0}\ket{r^z_i,r^x_i \oplus 1}, \label{eq:action_x} \\
    &\ZZ_{\langle i,j \rangle} \ket{r^z_i,r^x_i} \ket{r^z_j,r^x_j} = \delta_{r^x_i \oplus r^x_j ,0} \ket{r^z_i  \oplus 1,r^x_i} \ket{r^z_j  \oplus 1,r^x_j}, \label{eq:action_z}
\end{align}
where $\oplus$ denotes addition modulo 2.

In order to facilitate an efficient update scheme, we define the operators $\HH_{0,i} =  I_i \otimes I_i$, $\HH_{1,i} =  \XX_i$, $\HH_{2,\langle i,j \rangle} =  I_i I_j \otimes I_i I_j,$ and $\HH_{3,\langle i,j \rangle} =  \ZZ_{\langle i,j \rangle}$, such that $\HH = -2h\sum_{t=0,1} \sum_i \HH_{t,i} - 2\sum_{t=2,3} \sum_{\langle i,j \rangle} \HH_{t,{\langle i,j \rangle}} $ up to a constant energy shift. 
Following the standard SSE formulation~\cite{sandvik2010computational} and introducing a sufficiently large cut-off $M$ such that the truncation error is negligible, we reformulate the partition function as 
\begin{equation} 
    \Tilde{Z}_B = \sum_{n=0}^M \frac{(2\beta)^n (M-n)! }{M!} \sum_{S_M} \sum_{\{r^z, r^x\}}  \bra{r^z_0,r^x_0} \prod_{k=1}^M \HH_{b_k} \ket{r^z_0,r^x_0},
\end{equation}
where we additionally insert some null operators denoted by $\HH_{-1,-1}$. We can now sample the configurations, which include the Bell states and the operator list.
In the following, we will focus on the ground state ($\beta \to \infty$) simulations. Note that, with minimal modification, the algorithm can also be implemented in the projector QMC framework~\cite{sandvik2019sse,Melko2013sse}.

We can update the configurations using three types of updates. First, a diagonal update replaces the null operators by $\HH_{t\in \{0,2\},b}$, and vice versa. This update can be done as in the standard SSE~\cite{sandvik2010computational}. Second, a cluster update is used to change between $\HH_{0,i}$ and $\HH_{1,i}$, and is performed analogously to the cluster update in the TFIM~\cite{sandvik2003sse_ising}, while taking into account the constraints in Eq.~\eqref{eq:action_x} and \eqref{eq:action_z}. Finally, we introduce a bond cluster update, which changes between $\HH_{2,b}$ and $\HH_{3,b}$. This update can be seen as a cluster update that acts on the bond variables. More details on the update scheme can be found in the Supplemental Material~\cite{supmat}, which applies for general unfrustrated Ising interactions.

\newsec{Estimators}
We first consider the estimator for Pauli operators $\sigma_\mathbf{s}$, which can be written as a tensor product of single-qubit Pauli operators, namely $\sigma_\mathbf{s} = \otimes_i \sigma_{s^z_i s^x_i}$. The squared expectation value of any Pauli operators $\sigma_{\mathbf{s}}$ can be expressed as the expectation value of the operator $\sigma_{\mathbf{s}} \otimes \sigma_{\mathbf{s}}$, i.e. $\Tr[(\sigma_{\mathbf{s}} \otimes \sigma_{\mathbf{s}}) (\rho \otimes \rho)]=\Tr( \rho \sigma_{\mathbf{s}})^2$.
Since $\sigma_{\mathbf{s}} \otimes \sigma_{\mathbf{s}}$ is diagonal in the Bell basis, the estimator is straightforward: 
\begin{equation}
    \frac{\Tr( e^{-\beta H} \sigma_{\mathbf{s}})^2}{\Tr(e^{-\beta H})^2} = \left\langle \prod_i (-1)^{(s^x_i r^z_i - s^z_i r^x_i)} \right\rangle_{\mathbf{r}},
\end{equation}
where the average is over the sample of Bell states $\ket{\mathbf{r}^z,\mathbf{r}^x}$.
This is in sharp contrast with the conventional QMC in the computational basis, where only operators consisting of $Z$ operators are diagonal. %

Next, we discuss the estimator for the R\'enyi EE, which is defined as $S_n(A) = \frac{1}{1-n} \ln( \Tr \rho_A^n )$,
where $\rho_A = \Tr_{\Bar{A}} \rho$ is the reduced density matrix of a subsystem $A$. In the following, we will focus on $n=2$, in which case the purity can be written as $\Tr\rho_A^2 = \Tr (\mathbb{S}_A\rho \otimes \rho)$, where $\mathbb{S}_A=\otimes_{i\in A } \mathbb{S}_i$ and $\mathbb{S}$ is the SWAP operator which exchanges the two replicas, $\mathbb{S} \ket{a^{(1)}} \ket{a^{(2)}} = \ket{a^{(2)}} \ket{a^{(1)}}$. Importantly, the SWAP operator is also diagonal in the Bell basis, providing a simple estimator for $S_2$ as
\begin{equation} \label{eq:estimator_s2}
    S_2(A) = - \ln \left\langle \prod_{i\in A} (-1)^{r^x_i r^z_i} \right\rangle_{\mathbf{r}}.
\end{equation}
The estimation of the purity itself is always efficient in the number of samples. Indeed, the statistical error is quantified by the variance of the estimator, which is given by $\mathrm{Var}( \Tr\rho_A^2) = 1 - (\Tr\rho_A^2)^2\leq1$. As the variance is bounded, this ensures that the number of samples required to achieve a given additive error in the purity is constant. %
However, the R\'enyi-2 EE $S_2$ is the quantity of primary interest. By standard error propagation techniques, the variance for $S_2$ is given by $\mathrm{Var}(S_2) \approx (1 - (\Tr\rho_A^2)^2)/(\Tr\rho_A^2)^2 = e^{2S_2}-1$. In one-dimensional systems, where entanglement generally grows at most logarithmically with the subsystem size, the estimation of $S_2$ remains efficient, requiring only (at most) a polynomial number of samples. In higher-dimensional systems, where entanglement typically scales with the area of the subsystem, an exponentially growing number of samples is required to accurately estimate $S_2$. Importantly, this area-law scaling results in a much less severe exponential growth in sample complexity than the worst-case scenario of volume-law entanglement. This suggests a potential exponential gain in sample complexity over ratio estimators in conventional QMC.

\newsec{Thermodynamic integration}
As we will show below, the computation of entanglement using simple QMC measurements of Eq.~\eqref{eq:estimator_s2} already allows us to access large system sizes. However, the question arises: can we push these limits further? Previous methods relied on writing the entanglement as a ratio of partition functions, and applying techniques to mitigate the exponential sample problem. Here, it is crucial that both of the partition functions can be sampled from. At first glance, this appears to be inapplicable in the Bell-QMC framework due to the sign problem in the estimator in Eq.~\eqref{eq:estimator_s2}. We show in the End Matter that this limitation can be circumvented by employing an alternative estimator. The key idea is to introduce a partition function $Q(\lambda)$ for $0\leq \lambda \leq1$ such that $S_2=-\ln(Q(1)) + \ln(Q(0)) + N_A$. The difference between partition functions can then be evaluated using the thermodynamic integration (or related approach). Importantly, even in this scenario, Bell-QMC retains an advantage over conventional QMC, as the estimator can be simplified when working on the Bell basis, leading to significantly lower variances.

\newsec{Numerical results} \label{sec:num_results}
We now present our numerical results. We apply the Bell-QMC algorithm in 1D TFIM and $\Z_2$ LGT in 2D. The update schemes for both of the models as well as benchmarks at small systems are discussed in the Supplemental Material~\cite{supmat}. For each of the models, we set the temperature to be sufficiently low to simulate the ground state.

We first consider the 1D TFIM, with the Hamiltonian in Eq.~\eqref{eq:ising}, with open boundary condition. In this model, there is a symmetry-breaking transition from the ferromagnetic
phase to the paramagnetic phase at $h_c=1$, which is governed by the Ising CFT with central charge $c=1/2$~\cite{DiFrancesco}. 

\begin{figure}
\centering
\begin{overpic}[width=0.49\linewidth]{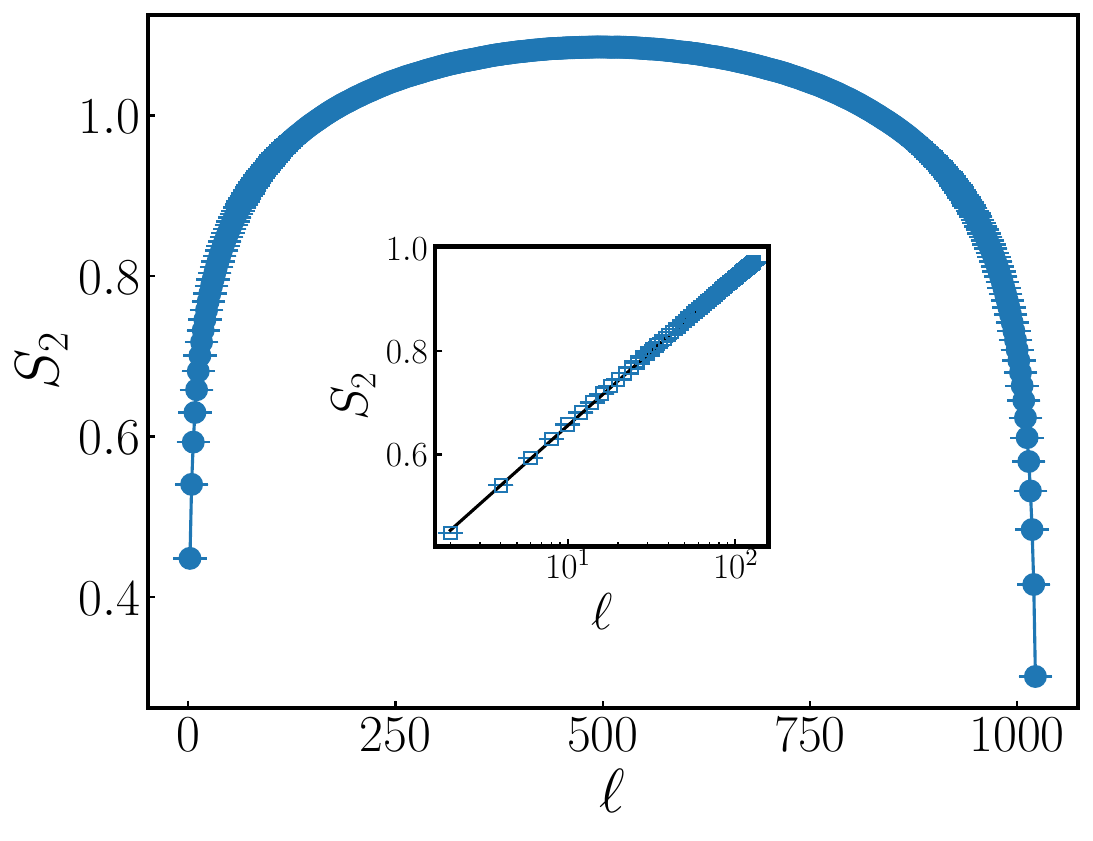}
\put (-3,70) {{\textbf{(a)}}}
\end{overpic}
\begin{overpic}[width=0.49\linewidth]{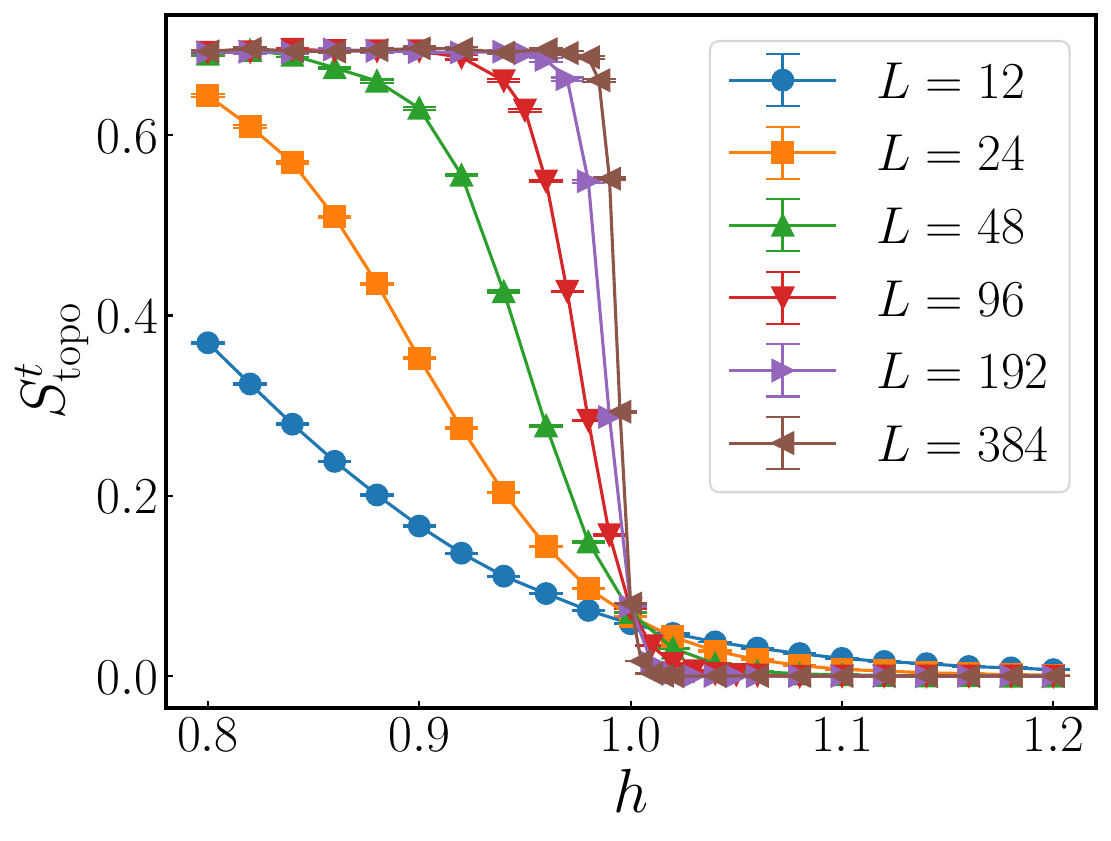}
\put (-3,70) {{\textbf{(b)}}}
\end{overpic}
\caption{ (a) R\'enyi-2 EE $S_2$ in the ground state of the TFIM at the critical point $h=1$ for partitions of sizes $\ell=\{2,4,\dots, 1022\}$ in the middle of the chain, with $L=1024$ and $\beta=3L$. Inset: logarithmic scaling of $S_2$ up to subsystem size $\ell=128$. The solid line denotes a linear fit, from which we extract the central charge $c \approx 0.5030(2)$. (b) The topological entanglement entropy $S^t_\mathrm{topo}$ as a function of the transverse field $h$ around the critical point $h=1$ and $\beta=4L$.} 
\label{fig:ising_mid_partitions}
\end{figure}

We will focus on the critical point $h_c$, where there is a logarithmic scaling of entanglement, as predicted by CFT~\cite{calabrese2004entanglement}. We first compute entanglement in the half bipartition for systems up to $L=1024$ and benchmark it against DMRG, obtaining excellent agreement as detailed in the Supplemental Material~\cite{supmat}. To emphasize the strength of Bell-QMC, we now consider subsystems in the middle of the chain, where computation becomes more involved in DMRG. Specifically, we consider the partitions $\{L/2-\ell/2+1, \dots, L/2 + \ell/2 \}$ of size $\ell$ (assuming even $\ell$). We recall that all these partitions can be estimated in a single Bell-QMC simulation. The results are shown in Fig.~\ref{fig:ising_mid_partitions}(a), showing clear logarithmic scaling, as expected from CFT. By fitting with the CFT scaling, $S_2 = \frac{c}{4} \ln \ell + \gamma$~\cite{calabrese2004entanglement}, we extract $c \approx 0.5030(2)$, which is consistent with the theoretical prediction $c=1/2$. %

In order to appreciate these results, it is worth to compare with earlier QMC studies. The first QMC algorithm to measure EE, introduced in Ref.~\cite{hastings2010measuring}, was formulated in the valence bond basis. While the EE can be estimated based on valence bond configurations, direct calculation using this estimator resulted in substantial statistical errors, even in the 1D Heisenberg model of $L= 100$ spins, thus motivating the introduction of the increment trick. Subsequently, Ref.~\cite{humeniuk2012quantum} proposed a general QMC scheme to calculate EE that can be implemented within SSE. Yet, even with the increment trick, the data quality remained poor for the 1D XX chain with $L=64$ spins. These examples illustrate that previous QMC techniques required intricate workarounds even in relatively simple one-dimensional systems. While further sophisticated tricks have been developed to alleviate these issues, they often involve technically complex and computationally intensive techniques. In stark contrast, our Bell-QMC method efficiently estimates the EE in one dimension using only diagonal measurements.

As an application, we investigate the symmetry-breaking transition in the TFIM from the lens of entanglement. In particular, we consider the (generalized) topological entanglement entropy, defined as $S^t_\mathrm{topo} =  S_2(AB) + S_2(BC) - S_2(ABC) - S_2(B)$,
which probe symmetry-breaking orders~\cite{zeng2015,zeng2019}. Here, the system is divided into three non-overlapping parts $A$, $B$, and $C$, arranged from left to right along an open chain. 
$S^t_\mathrm{topo}$ has proved to be a useful tool for identifying symmetry-breaking transitions, both in the context of ground states~\cite{zeng2015} and measurement-induced transitions~\cite{Lavasani2021,Klocke2022}. The results are shown in Fig. \ref{fig:ising_mid_partitions}(b), with $L=3L_A=3L_C=3L_B$. We see that $S^t_\mathrm{topo}$ takes a quantized value of $\ln{2}$ and $0$ in the symmetry-broken and paramagnetic phase, respectively, providing a large scale confirmation of the theoretical expectations~\cite{zeng2015}.

Next, we consider the $\Z_2$ lattice gauge theory in 2D, with Hamiltonian:
\begin{equation} \label{eq:z2_lgt}
    H = -\sum_\square \prod_{i\in \square} X_i - h\sum_i Z_i,
\end{equation}
where the spins live on the links of the square lattice. %
We are interested in the charge-free sector, that satisfies the Gauss' law $\prod_{i\in +} Z_i = 1$, on each vertices of the lattice. In fact, the ground state of Eq.~\eqref{eq:z2_lgt} is known to live in the charge-free sector, as can be established using perturbative arguments. In this sector, a deconfinement transition occurs at $h_c\approx 0.33$~\cite{blote2002cluster}. %

\begin{figure} 
\centering
\begin{overpic}[width=0.49\linewidth]{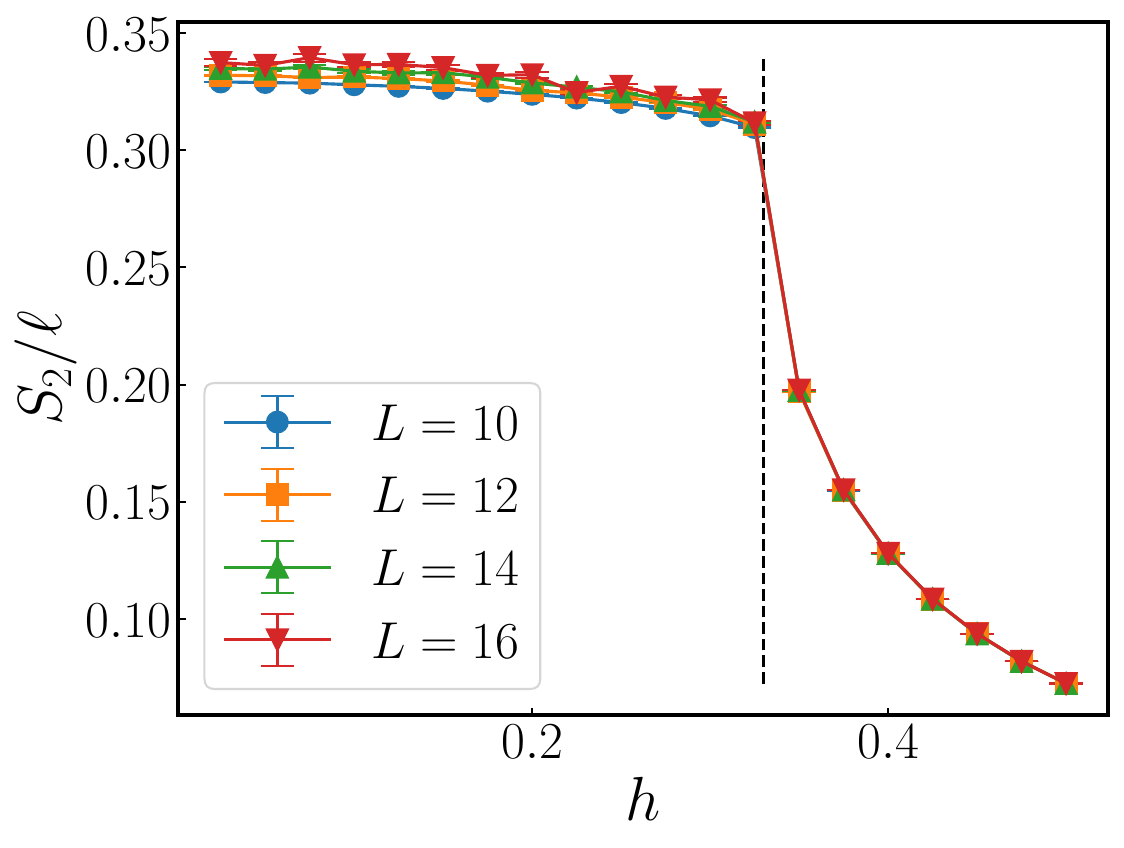}
\put (-3,70) {{\textbf{(a)}}}
\end{overpic}
\begin{overpic}[width=0.49\linewidth]{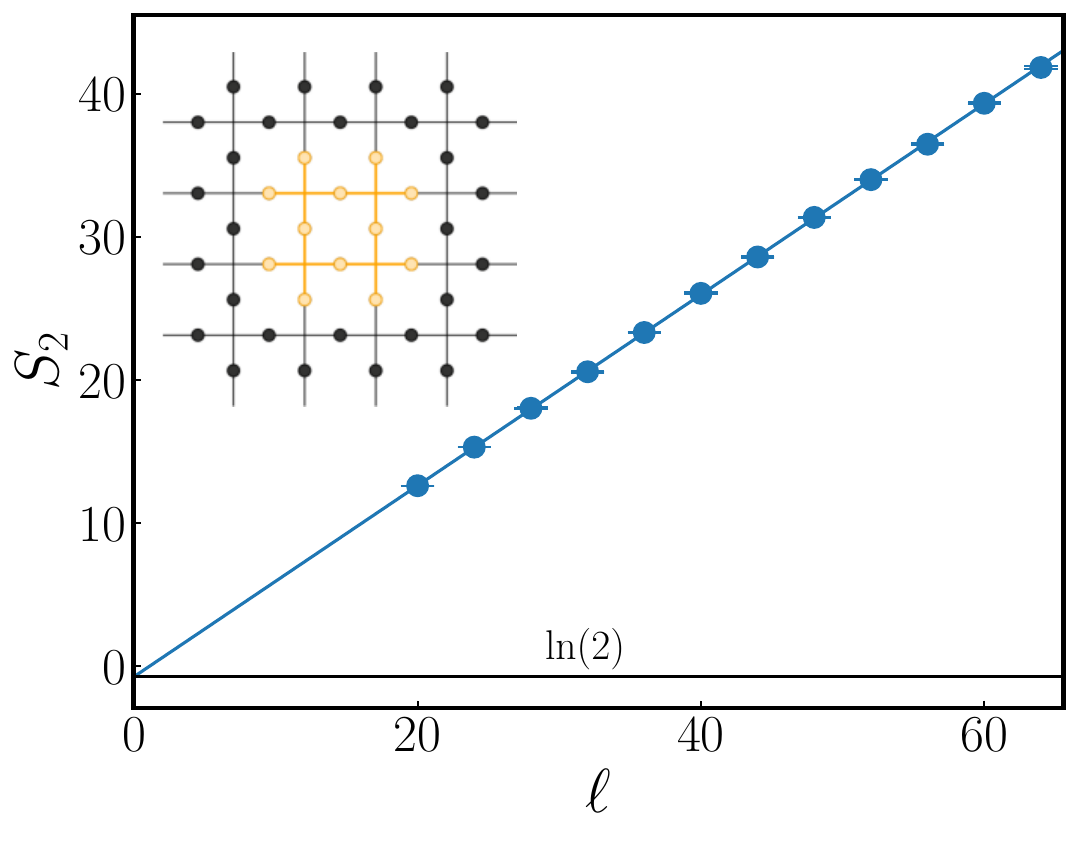}
\put (-3,70) {{\textbf{(b)}}}
\end{overpic}
\caption{R\'enyi-2 EE $S_2$ in the ground state of the $\Z_2$ lattice gauge theory. In (a), we show $S_2$ as a function of the transverse field $h$ around the critical point $h_c \approx 0.33$ for various system sizes with $\beta=4L$. In (b), we set $h=0.3$ and $\beta=L$, extrapolating $S_2$ to obtain the TEE $\gamma=0.71(3)$. In both plots, we consider a square partition of size $L/2 \times L/2$ with boundary length $\ell=2L$.}
\label{fig:z2_lgt_s2}
\end{figure}

We consider the entanglement of a square partition (see inset of Fig.~\ref{fig:z2_lgt_s2}b). In particular, for a torus of size $L\times L$, we consider $L/2 \times L/2$ partition, where the boundary length is $\ell=2L$. The gauge symmetry can be exploited to improve the estimator for $S_2$, as detailed in the Supplemental Material~\cite{supmat}.  Fig.~\ref{fig:z2_lgt_s2}a shows that $S_2$ satisfies area law at any $h$, and it monotonically decreases with increasing transverse field $h$. Moreover, there is an apparent non-analytic behavior around the critical point $h_c \approx 0.33$.

Further, we focus on the point $h=0.3 < h_c$ in the deconfined phase. The entanglement has the scaling form $S_2 = a\ell - \gamma$, where $\gamma$ is the topological entanglement entropy (TEE)~\cite{kitaev2006,levin2006}. %
As we show in Fig.~\ref{fig:z2_lgt_s2}b, the extrapolation of $S_2$ at $h=0.3$ yields $\gamma=0.71(3)$, in close agreement with the expected value $\ln 2$ for $\Z_2$ topological order. For this calculation, we employed the thermodynamic integration method for $L\geq22$, while smaller system sizes were calculated using direct estimation of Eq.~\eqref{eq:estimator_s2}. Remarkably, the data obtained solely from direct estimation (up to $L=20$) also yields a consistent value of  $\gamma=0.71(4)$.

Extracting the TEE from QMC simulations has long been an extremely challenging task, mainly due to the high computational cost of entanglement measurement. Despite intense research efforts, accurate TEE extraction in topologically ordered phases has only been achieved recently~\cite{zhao2022measuring}, utilizing computationally intensive method. Our Bell-QMC algorithm, using only simple diagonal measurements, is able to access larger system sizes than Ref.~\cite{zhao2022measuring} for the TEE calculation, highlighting the power of our algorithm. %

\begin{figure}
\centering
\begin{overpic}[width=0.49\linewidth]{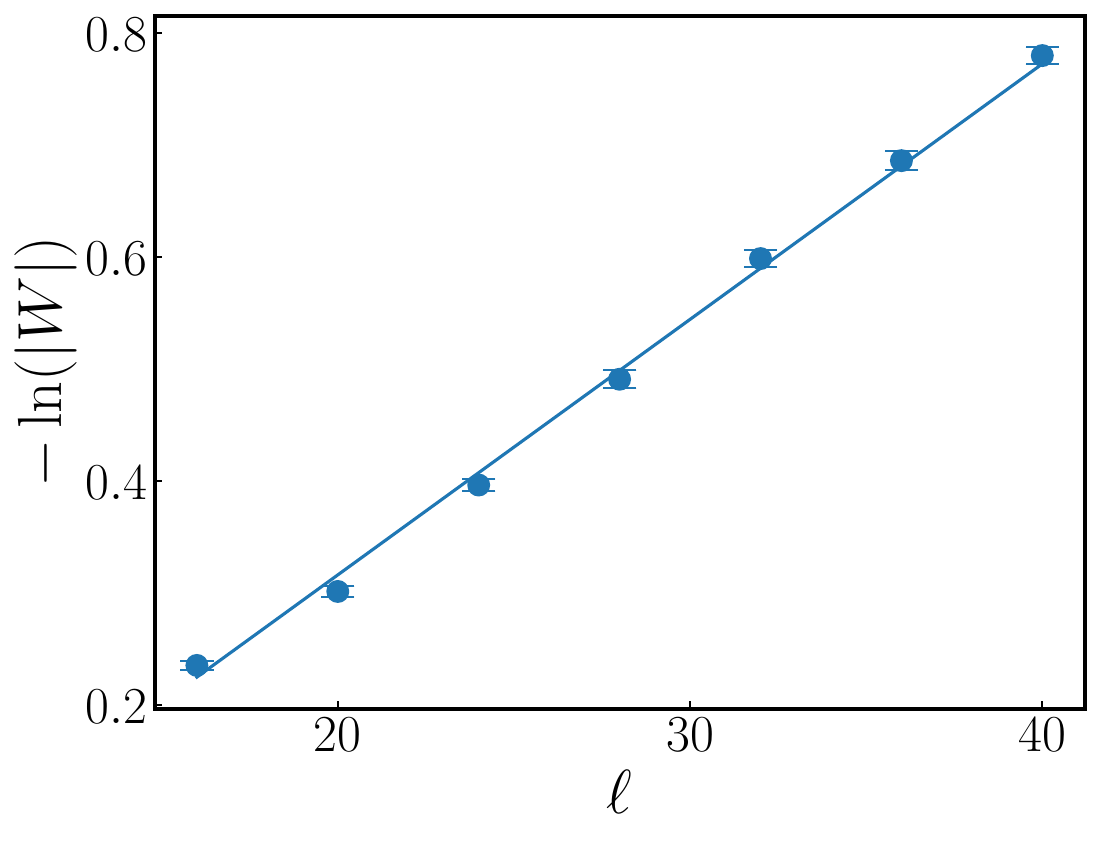}
\put (-3,70) {{\textbf{(a)}}}
\end{overpic}
\begin{overpic}[width=0.49\linewidth]{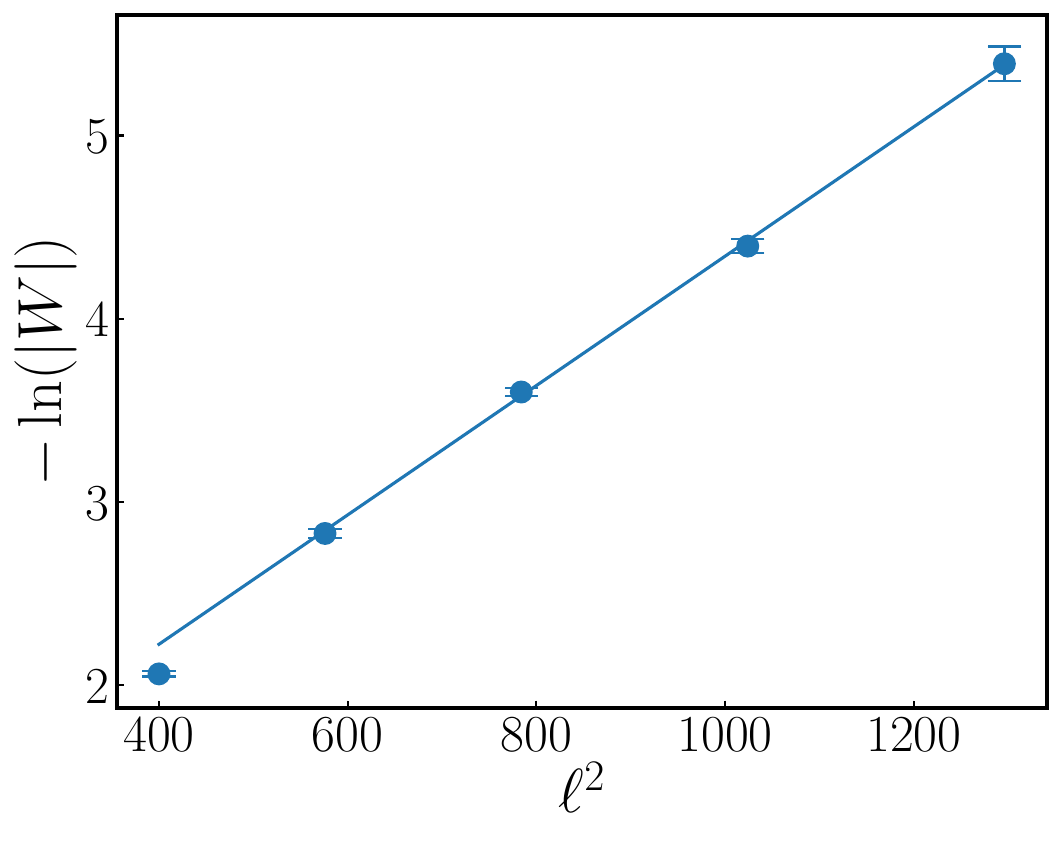}
\put (-3,70) {{\textbf{(b)}}}
\end{overpic}
\caption{The scaling of Wilson loop in the ground state of the $\Z_2$ lattice gauge theory at (a) $h=0.3$ in the deconfined phase and (b) $h=0.35$ in the confined phase. In both plots, we consider a square partition of size $L/2 \times L/2$ with boundary length $\ell=2L$ and temperature $\beta=4L$. 
The solid line denotes a linear fit.} 
\label{fig:z2_lgt_wilson}
\end{figure}

Finally, we consider the Wilson loop operators $W(C) = \prod_{i \in \partial C} X_i$, which creates an electric loop along a closed path $\partial C$, the boundary of region $C$ on the lattice. The Wilson loop operators characterize the confinement transition in $\Z_2$ gauge theory, displaying perimeter law in the deconfined phase and area law in the confined phase. The perimeter law corresponds to $\langle \lvert W \rvert \rangle \propto e^{-a \ell}$ while the area law corresponds to $\langle \lvert W \rvert \rangle \propto e^{-b \ell^2}$, where $\ell$ is the boundary length. Fig.~\ref{fig:z2_lgt_wilson} shows the scaling of Wilson loops at $h=0.3<h_c$ and $h=0.35>h_c$, confirming the expected scaling behavior.

\newsec{Conclusions and outlook} \label{sec:concl}
We have proposed a novel QMC framework that built upon the powerful quantum information protocol of Bell sampling. We discussed how this framework solves long-standing challenges in conventional QMC simulations, as it enables the unbiased calculation of all Pauli operators and the Rényi-2 entanglement entropy via straightforward diagonal measurements, 
with the same efficiency and simplicity as conventional QMC in computing, e.g., magnetization.
We have detailed the implementation of our algorithm within the SSE framework, including an efficient update scheme for general unfrustrated Ising interactions. %
Applying our algorithm in several scenarios, we have demonstrated how it surpasses state-of-the-art QMC simulations, all while retaining the conceptual simplicity of the algorithm.

Our work opens up many interesting directions. We expect that Bell-QMC would be useful to calculate nonlinear observables, crucial for the study of recently investigated phenomena such as measurement-induced phenomena~\cite{garratt2023measurements, lee2023quantum,murciano2023measurement,weinstein2023nonlocality,yang2023entanglemet,hoshino2024entanglementswapping,baweja2024postmeasurement} and the strong-to-weak spontaneous symmetry breaking phenomena~\cite{lessa2025strongtoweak,sala2024spontaneous}. Furthermore, another crucial area of exploration lies in the study of quantum resources within many-body quantum systems~\cite{chitambar2019quantum}. Beyond entanglement, nonstabilizerness, quantified by the Stabilizer Rényi Entropy (SRE)~\cite{leone2022stabilizer}, has garnered significant recent attention~\cite{haug2022quantifying,lami2023nonstabilizerness,haug2023stabilizer,tarabunga2023many,tarabunga2024nonstabilizernessmps,tarabunga2024critical,tarabunga2023magic,falcao2024nonstabilizerness}. It has been demonstrated that the SRE can be efficiently measured via numerical methods based on Pauli sampling~\cite{tarabunga2025efficient,lami2023nonstabilizerness,haug2023stabilizer,tarabunga2023many,collura2024quantum}, which directly suggests the application of Bell-QMC for measuring the SRE. It would be interesting to compare this approach to the recent QMC calculations of the SRE~\cite{liu2025nonequilibrium,ding2025evaluating}. %
Moreover, combined with stochastic analytic continuation methods~\cite{Sandvik1998sac,Syljuasen2008sac,Sandvik2016sac,Shao2023sac}, Bell-QMC can also obtain the spectral functions of general Pauli string operators, thereby enabling the investigation of a broader range of excitations.

From a methodological perspective, a natural next step is to extend the Bell-QMC approach to Heisenberg interactions, as well as other flavors of QMC. Finally, an interesting extension of our work would be to generalize the Bell-QMC method to measurements involving more than two copies of the state. This is likely essential for enabling the calculation of Rényi negativity~\cite{calabrese2012entanglement,ding2024negativity,wukaihsin2020negativity,wangfohong2025negativity,chung2014entanglement} and entanglement witnesses based on partial transpose moments~\cite{Elben2020pkppt,Neven2021pkppt,Yu2021,tarabunga2025quantifying}, which are crucial for characterizing and detecting entanglement in mixed states. 

\newsec{Acknowledgments}
We thank Z. Yan for helpful discussions.
 P.S.T. acknowledges funding by the Deutsche Forschungsgemeinschaft (DFG, German Research Foundation) under Germany’s Excellence Strategy – EXC-2111 – 390814868.

 \textbf{Data availability.}
Data and codes are available upon reasonable request on Zenodo~\cite{zenodo}.

\bibliographystyle{apsrev4-1}
\bibliography{bibliography}

\begin{thebibliography}{92}%
\makeatletter
\providecommand \@ifxundefined [1]{%
 \@ifx{#1\undefined}
}%
\providecommand \@ifnum [1]{%
 \ifnum #1\expandafter \@firstoftwo
 \else \expandafter \@secondoftwo
 \fi
}%
\providecommand \@ifx [1]{%
 \ifx #1\expandafter \@firstoftwo
 \else \expandafter \@secondoftwo
 \fi
}%
\providecommand \natexlab [1]{#1}%
\providecommand \enquote  [1]{``#1''}%
\providecommand \bibnamefont  [1]{#1}%
\providecommand \bibfnamefont [1]{#1}%
\providecommand \citenamefont [1]{#1}%
\providecommand \href@noop [0]{\@secondoftwo}%
\providecommand \href [0]{\begingroup \@sanitize@url \@href}%
\providecommand \@href[1]{\@@startlink{#1}\@@href}%
\providecommand \@@href[1]{\endgroup#1\@@endlink}%
\providecommand \@sanitize@url [0]{\catcode `\\12\catcode `\$12\catcode `\&12\catcode `\#12\catcode `\^12\catcode `\_12\catcode `\%12\relax}%
\providecommand \@@startlink[1]{}%
\providecommand \@@endlink[0]{}%
\providecommand \url  [0]{\begingroup\@sanitize@url \@url }%
\providecommand \@url [1]{\endgroup\@href {#1}{\urlprefix }}%
\providecommand \urlprefix  [0]{URL }%
\providecommand \Eprint [0]{\href }%
\providecommand \doibase [0]{http://dx.doi.org/}%
\providecommand \selectlanguage [0]{\@gobble}%
\providecommand \bibinfo  [0]{\@secondoftwo}%
\providecommand \bibfield  [0]{\@secondoftwo}%
\providecommand \translation [1]{[#1]}%
\providecommand \BibitemOpen [0]{}%
\providecommand \bibitemStop [0]{}%
\providecommand \bibitemNoStop [0]{.\EOS\space}%
\providecommand \EOS [0]{\spacefactor3000\relax}%
\providecommand \BibitemShut  [1]{\csname bibitem#1\endcsname}%
\let\auto@bib@innerbib\@empty
\bibitem [{\citenamefont {Schollwöck}(2011)}]{Schollwock2011dmrg}%
  \BibitemOpen
  \bibfield  {author} {\bibinfo {author} {\bibfnamefont {U.}~\bibnamefont {Schollwöck}},\ }\href {\doibase https://doi.org/10.1016/j.aop.2010.09.012} {\bibfield  {journal} {\bibinfo  {journal} {Annals of Physics}\ }\textbf {\bibinfo {volume} {326}},\ \bibinfo {pages} {96} (\bibinfo {year} {2011})},\ \bibinfo {note} {january 2011 Special Issue}\BibitemShut {NoStop}%
\bibitem [{\citenamefont {Orús}(2014)}]{orus2014practical}%
  \BibitemOpen
  \bibfield  {author} {\bibinfo {author} {\bibfnamefont {R.}~\bibnamefont {Orús}},\ }\href {\doibase https://doi.org/10.1016/j.aop.2014.06.013} {\bibfield  {journal} {\bibinfo  {journal} {Annals of Physics}\ }\textbf {\bibinfo {volume} {349}},\ \bibinfo {pages} {117} (\bibinfo {year} {2014})}\BibitemShut {NoStop}%
\bibitem [{\citenamefont {Sandvik}\ \emph {et~al.}(2010)\citenamefont {Sandvik}, \citenamefont {Avella},\ and\ \citenamefont {Mancini}}]{sandvik2010computational}%
  \BibitemOpen
  \bibfield  {author} {\bibinfo {author} {\bibfnamefont {A.~W.}\ \bibnamefont {Sandvik}}, \bibinfo {author} {\bibfnamefont {A.}~\bibnamefont {Avella}}, \ and\ \bibinfo {author} {\bibfnamefont {F.}~\bibnamefont {Mancini}},\ }in\ \href {\doibase 10.1063/1.3518900} {\emph {\bibinfo {booktitle} {AIP Conference Proceedings}}}\ (\bibinfo  {publisher} {AIP},\ \bibinfo {year} {2010})\BibitemShut {NoStop}%
\bibitem [{\citenamefont {Gubernatis}\ \emph {et~al.}(2016)\citenamefont {Gubernatis}, \citenamefont {Kawashima},\ and\ \citenamefont {Werner}}]{Gubernatis2016qmcbook}%
  \BibitemOpen
  \bibfield  {author} {\bibinfo {author} {\bibfnamefont {J.}~\bibnamefont {Gubernatis}}, \bibinfo {author} {\bibfnamefont {N.}~\bibnamefont {Kawashima}}, \ and\ \bibinfo {author} {\bibfnamefont {P.}~\bibnamefont {Werner}},\ }\href {\doibase 10.1017/CBO9780511902581} {\emph {\bibinfo {title} {Quantum Monte Carlo Methods: Algorithms for Lattice Models}}}\ (\bibinfo  {publisher} {Cambridge University Press},\ \bibinfo {year} {2016})\BibitemShut {NoStop}%
\bibitem [{\citenamefont {Becca}\ and\ \citenamefont {Sorella}(2017)}]{Becca2017}%
  \BibitemOpen
  \bibfield  {author} {\bibinfo {author} {\bibfnamefont {F.}~\bibnamefont {Becca}}\ and\ \bibinfo {author} {\bibfnamefont {S.}~\bibnamefont {Sorella}},\ }\href {\doibase 10.1017/9781316417041} {\emph {\bibinfo {title} {Quantum Monte Carlo Approaches for Correlated Systems}}}\ (\bibinfo  {publisher} {Cambridge University Press},\ \bibinfo {year} {2017})\BibitemShut {NoStop}%
\bibitem [{\citenamefont {Wolf}\ \emph {et~al.}(2008)\citenamefont {Wolf}, \citenamefont {Verstraete}, \citenamefont {Hastings},\ and\ \citenamefont {Cirac}}]{wolf2008area}%
  \BibitemOpen
  \bibfield  {author} {\bibinfo {author} {\bibfnamefont {M.~M.}\ \bibnamefont {Wolf}}, \bibinfo {author} {\bibfnamefont {F.}~\bibnamefont {Verstraete}}, \bibinfo {author} {\bibfnamefont {M.~B.}\ \bibnamefont {Hastings}}, \ and\ \bibinfo {author} {\bibfnamefont {J.~I.}\ \bibnamefont {Cirac}},\ }\href {\doibase 10.1103/PhysRevLett.100.070502} {\bibfield  {journal} {\bibinfo  {journal} {Phys. Rev. Lett.}\ }\textbf {\bibinfo {volume} {100}},\ \bibinfo {pages} {070502} (\bibinfo {year} {2008})}\BibitemShut {NoStop}%
\bibitem [{\citenamefont {Calabrese}\ and\ \citenamefont {Cardy}(2004)}]{calabrese2004entanglement}%
  \BibitemOpen
  \bibfield  {author} {\bibinfo {author} {\bibfnamefont {P.}~\bibnamefont {Calabrese}}\ and\ \bibinfo {author} {\bibfnamefont {J.}~\bibnamefont {Cardy}},\ }\href {\doibase 10.1088/1742-5468/2004/06/p06002} {\bibfield  {journal} {\bibinfo  {journal} {Journal of Statistical Mechanics: Theory and Experiment}\ }\textbf {\bibinfo {volume} {2004}},\ \bibinfo {pages} {P06002} (\bibinfo {year} {2004})}\BibitemShut {NoStop}%
\bibitem [{\citenamefont {Metlitski}\ and\ \citenamefont {Grover}(2015)}]{metlitski2015entanglement}%
  \BibitemOpen
  \bibfield  {author} {\bibinfo {author} {\bibfnamefont {M.~A.}\ \bibnamefont {Metlitski}}\ and\ \bibinfo {author} {\bibfnamefont {T.}~\bibnamefont {Grover}},\ }\href {https://arxiv.org/abs/1112.5166} {\bibfield  {journal} {\bibinfo  {journal} {arxiv:1112.5166}\ } (\bibinfo {year} {2015})}\BibitemShut {NoStop}%
\bibitem [{\citenamefont {Kitaev}\ and\ \citenamefont {Preskill}(2006)}]{kitaev2006}%
  \BibitemOpen
  \bibfield  {author} {\bibinfo {author} {\bibfnamefont {A.}~\bibnamefont {Kitaev}}\ and\ \bibinfo {author} {\bibfnamefont {J.}~\bibnamefont {Preskill}},\ }\href {\doibase 10.1103/PhysRevLett.96.110404} {\bibfield  {journal} {\bibinfo  {journal} {Phys. Rev. Lett.}\ }\textbf {\bibinfo {volume} {96}},\ \bibinfo {pages} {110404} (\bibinfo {year} {2006})}\BibitemShut {NoStop}%
\bibitem [{\citenamefont {Levin}\ and\ \citenamefont {Wen}(2006)}]{levin2006}%
  \BibitemOpen
  \bibfield  {author} {\bibinfo {author} {\bibfnamefont {M.}~\bibnamefont {Levin}}\ and\ \bibinfo {author} {\bibfnamefont {X.-G.}\ \bibnamefont {Wen}},\ }\href {\doibase 10.1103/PhysRevLett.96.110405} {\bibfield  {journal} {\bibinfo  {journal} {Phys. Rev. Lett.}\ }\textbf {\bibinfo {volume} {96}},\ \bibinfo {pages} {110405} (\bibinfo {year} {2006})}\BibitemShut {NoStop}%
\bibitem [{\citenamefont {Amico}\ \emph {et~al.}(2008)\citenamefont {Amico}, \citenamefont {Fazio}, \citenamefont {Osterloh},\ and\ \citenamefont {Vedral}}]{amico2008}%
  \BibitemOpen
  \bibfield  {author} {\bibinfo {author} {\bibfnamefont {L.}~\bibnamefont {Amico}}, \bibinfo {author} {\bibfnamefont {R.}~\bibnamefont {Fazio}}, \bibinfo {author} {\bibfnamefont {A.}~\bibnamefont {Osterloh}}, \ and\ \bibinfo {author} {\bibfnamefont {V.}~\bibnamefont {Vedral}},\ }\href {\doibase 10.1103/RevModPhys.80.517} {\bibfield  {journal} {\bibinfo  {journal} {Rev. Mod. Phys.}\ }\textbf {\bibinfo {volume} {80}},\ \bibinfo {pages} {517} (\bibinfo {year} {2008})}\BibitemShut {NoStop}%
\bibitem [{\citenamefont {Laflorencie}(2016)}]{Laflorencie2016entReview}%
  \BibitemOpen
  \bibfield  {author} {\bibinfo {author} {\bibfnamefont {N.}~\bibnamefont {Laflorencie}},\ }\href {\doibase https://doi.org/10.1016/j.physrep.2016.06.008} {\bibfield  {journal} {\bibinfo  {journal} {Physics Reports}\ }\textbf {\bibinfo {volume} {646}},\ \bibinfo {pages} {1} (\bibinfo {year} {2016})}\BibitemShut {NoStop}%
\bibitem [{\citenamefont {Hastings}\ \emph {et~al.}(2010)\citenamefont {Hastings}, \citenamefont {González}, \citenamefont {Kallin},\ and\ \citenamefont {Melko}}]{hastings2010measuring}%
  \BibitemOpen
  \bibfield  {author} {\bibinfo {author} {\bibfnamefont {M.~B.}\ \bibnamefont {Hastings}}, \bibinfo {author} {\bibfnamefont {I.}~\bibnamefont {González}}, \bibinfo {author} {\bibfnamefont {A.~B.}\ \bibnamefont {Kallin}}, \ and\ \bibinfo {author} {\bibfnamefont {R.~G.}\ \bibnamefont {Melko}},\ }\href {\doibase 10.1103/physrevlett.104.157201} {\bibfield  {journal} {\bibinfo  {journal} {Physical Review Letters}\ }\textbf {\bibinfo {volume} {104}} (\bibinfo {year} {2010}),\ 10.1103/physrevlett.104.157201}\BibitemShut {NoStop}%
\bibitem [{\citenamefont {Humeniuk}\ and\ \citenamefont {Roscilde}(2012)}]{humeniuk2012quantum}%
  \BibitemOpen
  \bibfield  {author} {\bibinfo {author} {\bibfnamefont {S.}~\bibnamefont {Humeniuk}}\ and\ \bibinfo {author} {\bibfnamefont {T.}~\bibnamefont {Roscilde}},\ }\href {\doibase 10.1103/physrevb.86.235116} {\bibfield  {journal} {\bibinfo  {journal} {Physical Review B}\ }\textbf {\bibinfo {volume} {86}} (\bibinfo {year} {2012}),\ 10.1103/physrevb.86.235116}\BibitemShut {NoStop}%
\bibitem [{\citenamefont {Zhou}\ \emph {et~al.}(2024)\citenamefont {Zhou}, \citenamefont {Meng}, \citenamefont {Qi},\ and\ \citenamefont {Da~Liao}}]{ZhouX2024incremental}%
  \BibitemOpen
  \bibfield  {author} {\bibinfo {author} {\bibfnamefont {X.}~\bibnamefont {Zhou}}, \bibinfo {author} {\bibfnamefont {Z.~Y.}\ \bibnamefont {Meng}}, \bibinfo {author} {\bibfnamefont {Y.}~\bibnamefont {Qi}}, \ and\ \bibinfo {author} {\bibfnamefont {Y.}~\bibnamefont {Da~Liao}},\ }\href {\doibase 10.1103/PhysRevB.109.165106} {\bibfield  {journal} {\bibinfo  {journal} {Phys. Rev. B}\ }\textbf {\bibinfo {volume} {109}},\ \bibinfo {pages} {165106} (\bibinfo {year} {2024})}\BibitemShut {NoStop}%
\bibitem [{\citenamefont {Melko}\ \emph {et~al.}(2010)\citenamefont {Melko}, \citenamefont {Kallin},\ and\ \citenamefont {Hastings}}]{melko2010finitesizescaling}%
  \BibitemOpen
  \bibfield  {author} {\bibinfo {author} {\bibfnamefont {R.~G.}\ \bibnamefont {Melko}}, \bibinfo {author} {\bibfnamefont {A.~B.}\ \bibnamefont {Kallin}}, \ and\ \bibinfo {author} {\bibfnamefont {M.~B.}\ \bibnamefont {Hastings}},\ }\href {\doibase 10.1103/physrevb.82.100409} {\bibfield  {journal} {\bibinfo  {journal} {Physical Review B}\ }\textbf {\bibinfo {volume} {82}} (\bibinfo {year} {2010}),\ 10.1103/physrevb.82.100409}\BibitemShut {NoStop}%
\bibitem [{\citenamefont {Ding}\ \emph {et~al.}(2024{\natexlab{a}})\citenamefont {Ding}, \citenamefont {Sun}, \citenamefont {Ma}, \citenamefont {Pan}, \citenamefont {Cheng},\ and\ \citenamefont {Yan}}]{ding2024reweight}%
  \BibitemOpen
  \bibfield  {author} {\bibinfo {author} {\bibfnamefont {Y.-M.}\ \bibnamefont {Ding}}, \bibinfo {author} {\bibfnamefont {J.-S.}\ \bibnamefont {Sun}}, \bibinfo {author} {\bibfnamefont {N.}~\bibnamefont {Ma}}, \bibinfo {author} {\bibfnamefont {G.}~\bibnamefont {Pan}}, \bibinfo {author} {\bibfnamefont {C.}~\bibnamefont {Cheng}}, \ and\ \bibinfo {author} {\bibfnamefont {Z.}~\bibnamefont {Yan}},\ }\href {\doibase 10.1103/physrevb.110.165152} {\bibfield  {journal} {\bibinfo  {journal} {Physical Review B}\ }\textbf {\bibinfo {volume} {110}} (\bibinfo {year} {2024}{\natexlab{a}}),\ 10.1103/physrevb.110.165152}\BibitemShut {NoStop}%
\bibitem [{\citenamefont {Wang}\ \emph {et~al.}(2025)\citenamefont {Wang}, \citenamefont {Wang}, \citenamefont {Ding}, \citenamefont {Mao},\ and\ \citenamefont {Yan}}]{Wang2024ent}%
  \BibitemOpen
  \bibfield  {author} {\bibinfo {author} {\bibfnamefont {Z.}~\bibnamefont {Wang}}, \bibinfo {author} {\bibfnamefont {Z.}~\bibnamefont {Wang}}, \bibinfo {author} {\bibfnamefont {Y.-M.}\ \bibnamefont {Ding}}, \bibinfo {author} {\bibfnamefont {B.-B.}\ \bibnamefont {Mao}}, \ and\ \bibinfo {author} {\bibfnamefont {Z.}~\bibnamefont {Yan}},\ }\href {\doibase 10.1038/s41467-025-61084-7} {\bibfield  {journal} {\bibinfo  {journal} {Nature Communications}\ }\textbf {\bibinfo {volume} {16}} (\bibinfo {year} {2025}),\ 10.1038/s41467-025-61084-7}\BibitemShut {NoStop}%
\bibitem [{\citenamefont {D’Emidio}(2020)}]{emidio2020entanglement}%
  \BibitemOpen
  \bibfield  {author} {\bibinfo {author} {\bibfnamefont {J.}~\bibnamefont {D’Emidio}},\ }\href {\doibase 10.1103/physrevlett.124.110602} {\bibfield  {journal} {\bibinfo  {journal} {Physical Review Letters}\ }\textbf {\bibinfo {volume} {124}} (\bibinfo {year} {2020}),\ 10.1103/physrevlett.124.110602}\BibitemShut {NoStop}%
\bibitem [{\citenamefont {Wang}\ \emph {et~al.}(2024)\citenamefont {Wang}, \citenamefont {Liu}, \citenamefont {Mao}, \citenamefont {Wang},\ and\ \citenamefont {Yan}}]{zhiyan2024offdiag}%
  \BibitemOpen
  \bibfield  {author} {\bibinfo {author} {\bibfnamefont {Z.}~\bibnamefont {Wang}}, \bibinfo {author} {\bibfnamefont {Z.}~\bibnamefont {Liu}}, \bibinfo {author} {\bibfnamefont {B.-B.}\ \bibnamefont {Mao}}, \bibinfo {author} {\bibfnamefont {Z.}~\bibnamefont {Wang}}, \ and\ \bibinfo {author} {\bibfnamefont {Z.}~\bibnamefont {Yan}},\ }\href {https://arxiv.org/abs/2412.01384} {\bibfield  {journal} {\bibinfo  {journal} {arXiv:2412.01384}\ } (\bibinfo {year} {2024})}\BibitemShut {NoStop}%
\bibitem [{\citenamefont {Hangleiter}\ and\ \citenamefont {Gullans}(2024)}]{hangleiter2024bell}%
  \BibitemOpen
  \bibfield  {author} {\bibinfo {author} {\bibfnamefont {D.}~\bibnamefont {Hangleiter}}\ and\ \bibinfo {author} {\bibfnamefont {M.~J.}\ \bibnamefont {Gullans}},\ }\href {\doibase 10.1103/physrevlett.133.020601} {\bibfield  {journal} {\bibinfo  {journal} {Physical Review Letters}\ }\textbf {\bibinfo {volume} {133}} (\bibinfo {year} {2024}),\ 10.1103/physrevlett.133.020601}\BibitemShut {NoStop}%
\bibitem [{\citenamefont {Huang}\ \emph {et~al.}(2022)\citenamefont {Huang}, \citenamefont {Broughton}, \citenamefont {Cotler}, \citenamefont {Chen}, \citenamefont {Li}, \citenamefont {Mohseni}, \citenamefont {Neven}, \citenamefont {Babbush}, \citenamefont {Kueng}, \citenamefont {Preskill},\ and\ \citenamefont {McClean}}]{huang2022quantum}%
  \BibitemOpen
  \bibfield  {author} {\bibinfo {author} {\bibfnamefont {H.-Y.}\ \bibnamefont {Huang}}, \bibinfo {author} {\bibfnamefont {M.}~\bibnamefont {Broughton}}, \bibinfo {author} {\bibfnamefont {J.}~\bibnamefont {Cotler}}, \bibinfo {author} {\bibfnamefont {S.}~\bibnamefont {Chen}}, \bibinfo {author} {\bibfnamefont {J.}~\bibnamefont {Li}}, \bibinfo {author} {\bibfnamefont {M.}~\bibnamefont {Mohseni}}, \bibinfo {author} {\bibfnamefont {H.}~\bibnamefont {Neven}}, \bibinfo {author} {\bibfnamefont {R.}~\bibnamefont {Babbush}}, \bibinfo {author} {\bibfnamefont {R.}~\bibnamefont {Kueng}}, \bibinfo {author} {\bibfnamefont {J.}~\bibnamefont {Preskill}}, \ and\ \bibinfo {author} {\bibfnamefont {J.~R.}\ \bibnamefont {McClean}},\ }\href {\doibase 10.1126/science.abn7293} {\bibfield  {journal} {\bibinfo  {journal} {Science}\ }\textbf {\bibinfo {volume} {376}},\ \bibinfo {pages} {1182–1186} (\bibinfo {year} {2022})}\BibitemShut {NoStop}%
\bibitem [{\citenamefont {Huang}\ \emph {et~al.}(2021)\citenamefont {Huang}, \citenamefont {Kueng},\ and\ \citenamefont {Preskill}}]{huang2021information}%
  \BibitemOpen
  \bibfield  {author} {\bibinfo {author} {\bibfnamefont {H.-Y.}\ \bibnamefont {Huang}}, \bibinfo {author} {\bibfnamefont {R.}~\bibnamefont {Kueng}}, \ and\ \bibinfo {author} {\bibfnamefont {J.}~\bibnamefont {Preskill}},\ }\href {\doibase 10.1103/physrevlett.126.190505} {\bibfield  {journal} {\bibinfo  {journal} {Physical Review Letters}\ }\textbf {\bibinfo {volume} {126}} (\bibinfo {year} {2021}),\ 10.1103/physrevlett.126.190505}\BibitemShut {NoStop}%
\bibitem [{\citenamefont {Haug}\ and\ \citenamefont {Kim}(2023)}]{haug2023scalable}%
  \BibitemOpen
  \bibfield  {author} {\bibinfo {author} {\bibfnamefont {T.}~\bibnamefont {Haug}}\ and\ \bibinfo {author} {\bibfnamefont {M.}~\bibnamefont {Kim}},\ }\href {\doibase 10.1103/prxquantum.4.010301} {\bibfield  {journal} {\bibinfo  {journal} {PRX Quantum}\ }\textbf {\bibinfo {volume} {4}} (\bibinfo {year} {2023}),\ 10.1103/prxquantum.4.010301}\BibitemShut {NoStop}%
\bibitem [{\citenamefont {Haug}\ \emph {et~al.}(2024{\natexlab{a}})\citenamefont {Haug}, \citenamefont {Lee},\ and\ \citenamefont {Kim}}]{haug2024efficient}%
  \BibitemOpen
  \bibfield  {author} {\bibinfo {author} {\bibfnamefont {T.}~\bibnamefont {Haug}}, \bibinfo {author} {\bibfnamefont {S.}~\bibnamefont {Lee}}, \ and\ \bibinfo {author} {\bibfnamefont {M.}~\bibnamefont {Kim}},\ }\href {\doibase 10.1103/physrevlett.132.240602} {\bibfield  {journal} {\bibinfo  {journal} {Physical Review Letters}\ }\textbf {\bibinfo {volume} {132}} (\bibinfo {year} {2024}{\natexlab{a}}),\ 10.1103/physrevlett.132.240602}\BibitemShut {NoStop}%
\bibitem [{\citenamefont {Montanaro}(2017)}]{montanaro2017learning}%
  \BibitemOpen
  \bibfield  {author} {\bibinfo {author} {\bibfnamefont {A.}~\bibnamefont {Montanaro}},\ }\href {https://arxiv.org/abs/1707.04012} {\bibfield  {journal} {\bibinfo  {journal} {arxiv:1707.04012}\ } (\bibinfo {year} {2017})}\BibitemShut {NoStop}%
\bibitem [{\citenamefont {Gross}\ \emph {et~al.}(2021)\citenamefont {Gross}, \citenamefont {Nezami},\ and\ \citenamefont {Walter}}]{gross2021schur}%
  \BibitemOpen
  \bibfield  {author} {\bibinfo {author} {\bibfnamefont {D.}~\bibnamefont {Gross}}, \bibinfo {author} {\bibfnamefont {S.}~\bibnamefont {Nezami}}, \ and\ \bibinfo {author} {\bibfnamefont {M.}~\bibnamefont {Walter}},\ }\href {\doibase 10.1007/s00220-021-04118-7} {\bibfield  {journal} {\bibinfo  {journal} {Communications in Mathematical Physics}\ }\textbf {\bibinfo {volume} {385}},\ \bibinfo {pages} {1325–1393} (\bibinfo {year} {2021})}\BibitemShut {NoStop}%
\bibitem [{\citenamefont {Grewal}\ \emph {et~al.}(2024)\citenamefont {Grewal}, \citenamefont {Iyer}, \citenamefont {Kretschmer},\ and\ \citenamefont {Liang}}]{grewal2024efficient}%
  \BibitemOpen
  \bibfield  {author} {\bibinfo {author} {\bibfnamefont {S.}~\bibnamefont {Grewal}}, \bibinfo {author} {\bibfnamefont {V.}~\bibnamefont {Iyer}}, \bibinfo {author} {\bibfnamefont {W.}~\bibnamefont {Kretschmer}}, \ and\ \bibinfo {author} {\bibfnamefont {D.}~\bibnamefont {Liang}},\ }\href {https://arxiv.org/abs/2305.13409} {\bibfield  {journal} {\bibinfo  {journal} {arxiv:2305.13409}\ } (\bibinfo {year} {2024})}\BibitemShut {NoStop}%
\bibitem [{\citenamefont {King}\ \emph {et~al.}(2025)\citenamefont {King}, \citenamefont {Gosset}, \citenamefont {Kothari},\ and\ \citenamefont {Babbush}}]{king2024triplyefficient}%
  \BibitemOpen
  \bibfield  {author} {\bibinfo {author} {\bibfnamefont {R.}~\bibnamefont {King}}, \bibinfo {author} {\bibfnamefont {D.}~\bibnamefont {Gosset}}, \bibinfo {author} {\bibfnamefont {R.}~\bibnamefont {Kothari}}, \ and\ \bibinfo {author} {\bibfnamefont {R.}~\bibnamefont {Babbush}},\ }\href {\doibase 10.1103/prxquantum.6.010336} {\bibfield  {journal} {\bibinfo  {journal} {PRX Quantum}\ }\textbf {\bibinfo {volume} {6}} (\bibinfo {year} {2025}),\ 10.1103/prxquantum.6.010336}\BibitemShut {NoStop}%
\bibitem [{\citenamefont {Chen}\ \emph {et~al.}(2022)\citenamefont {Chen}, \citenamefont {Cotler}, \citenamefont {Huang},\ and\ \citenamefont {Li}}]{chen2022exponential}%
  \BibitemOpen
  \bibfield  {author} {\bibinfo {author} {\bibfnamefont {S.}~\bibnamefont {Chen}}, \bibinfo {author} {\bibfnamefont {J.}~\bibnamefont {Cotler}}, \bibinfo {author} {\bibfnamefont {H.-Y.}\ \bibnamefont {Huang}}, \ and\ \bibinfo {author} {\bibfnamefont {J.}~\bibnamefont {Li}},\ }in\ \href {\doibase 10.1109/FOCS52979.2021.00063} {\emph {\bibinfo {booktitle} {2021 IEEE 62nd Annual Symposium on Foundations of Computer Science (FOCS)}}}\ (\bibinfo  {publisher} {IEEE Computer Society},\ \bibinfo {address} {Los Alamitos, CA, USA},\ \bibinfo {year} {2022})\ pp.\ \bibinfo {pages} {574--585}\BibitemShut {NoStop}%
\bibitem [{\citenamefont {Bluvstein}\ \emph {et~al.}(2024)\citenamefont {Bluvstein}, \citenamefont {Evered}, \citenamefont {Geim}, \citenamefont {Li}, \citenamefont {Zhou}, \citenamefont {Manovitz}, \citenamefont {Ebadi}, \citenamefont {Cain}, \citenamefont {Kalinowski}, \citenamefont {Hangleiter} \emph {et~al.}}]{bluvstein2024logical}%
  \BibitemOpen
  \bibfield  {author} {\bibinfo {author} {\bibfnamefont {D.}~\bibnamefont {Bluvstein}}, \bibinfo {author} {\bibfnamefont {S.~J.}\ \bibnamefont {Evered}}, \bibinfo {author} {\bibfnamefont {A.~A.}\ \bibnamefont {Geim}}, \bibinfo {author} {\bibfnamefont {S.~H.}\ \bibnamefont {Li}}, \bibinfo {author} {\bibfnamefont {H.}~\bibnamefont {Zhou}}, \bibinfo {author} {\bibfnamefont {T.}~\bibnamefont {Manovitz}}, \bibinfo {author} {\bibfnamefont {S.}~\bibnamefont {Ebadi}}, \bibinfo {author} {\bibfnamefont {M.}~\bibnamefont {Cain}}, \bibinfo {author} {\bibfnamefont {M.}~\bibnamefont {Kalinowski}}, \bibinfo {author} {\bibfnamefont {D.}~\bibnamefont {Hangleiter}},  \emph {et~al.},\ }\href {\doibase 10.1038/s41586-023-06927-3} {\bibfield  {journal} {\bibinfo  {journal} {Nature}\ }\textbf {\bibinfo {volume} {626}},\ \bibinfo {pages} {58} (\bibinfo {year} {2024})}\BibitemShut {NoStop}%
\bibitem [{\citenamefont {Haug}\ and\ \citenamefont {Tarabunga}(2025)}]{haug2025efficientwitnessingtestingmagic}%
  \BibitemOpen
  \bibfield  {author} {\bibinfo {author} {\bibfnamefont {T.}~\bibnamefont {Haug}}\ and\ \bibinfo {author} {\bibfnamefont {P.~S.}\ \bibnamefont {Tarabunga}},\ }\href {https://arxiv.org/abs/2504.18098} {\bibfield  {journal} {\bibinfo  {journal} {arxiv:2504.18098}\ } (\bibinfo {year} {2025})}\BibitemShut {NoStop}%
\bibitem [{\citenamefont {Islam}\ \emph {et~al.}(2015)\citenamefont {Islam}, \citenamefont {Ma}, \citenamefont {Preiss}, \citenamefont {Eric~Tai}, \citenamefont {Lukin}, \citenamefont {Rispoli},\ and\ \citenamefont {Greiner}}]{islam2015measuring}%
  \BibitemOpen
  \bibfield  {author} {\bibinfo {author} {\bibfnamefont {R.}~\bibnamefont {Islam}}, \bibinfo {author} {\bibfnamefont {R.}~\bibnamefont {Ma}}, \bibinfo {author} {\bibfnamefont {P.~M.}\ \bibnamefont {Preiss}}, \bibinfo {author} {\bibfnamefont {M.}~\bibnamefont {Eric~Tai}}, \bibinfo {author} {\bibfnamefont {A.}~\bibnamefont {Lukin}}, \bibinfo {author} {\bibfnamefont {M.}~\bibnamefont {Rispoli}}, \ and\ \bibinfo {author} {\bibfnamefont {M.}~\bibnamefont {Greiner}},\ }\href {\doibase 10.1038/nature15750} {\bibfield  {journal} {\bibinfo  {journal} {Nature}\ }\textbf {\bibinfo {volume} {528}},\ \bibinfo {pages} {77–83} (\bibinfo {year} {2015})}\BibitemShut {NoStop}%
\bibitem [{\citenamefont {Kaufman}\ \emph {et~al.}(2016)\citenamefont {Kaufman}, \citenamefont {Tai}, \citenamefont {Lukin}, \citenamefont {Rispoli}, \citenamefont {Schittko}, \citenamefont {Preiss},\ and\ \citenamefont {Greiner}}]{Kaufman2016}%
  \BibitemOpen
  \bibfield  {author} {\bibinfo {author} {\bibfnamefont {A.~M.}\ \bibnamefont {Kaufman}}, \bibinfo {author} {\bibfnamefont {M.~E.}\ \bibnamefont {Tai}}, \bibinfo {author} {\bibfnamefont {A.}~\bibnamefont {Lukin}}, \bibinfo {author} {\bibfnamefont {M.}~\bibnamefont {Rispoli}}, \bibinfo {author} {\bibfnamefont {R.}~\bibnamefont {Schittko}}, \bibinfo {author} {\bibfnamefont {P.~M.}\ \bibnamefont {Preiss}}, \ and\ \bibinfo {author} {\bibfnamefont {M.}~\bibnamefont {Greiner}},\ }\href {\doibase 10.1126/science.aaf6725} {\bibfield  {journal} {\bibinfo  {journal} {Science}\ }\textbf {\bibinfo {volume} {353}},\ \bibinfo {pages} {794–800} (\bibinfo {year} {2016})}\BibitemShut {NoStop}%
\bibitem [{\citenamefont {Linke}\ \emph {et~al.}(2018)\citenamefont {Linke}, \citenamefont {Johri}, \citenamefont {Figgatt}, \citenamefont {Landsman}, \citenamefont {Matsuura},\ and\ \citenamefont {Monroe}}]{Linke2018}%
  \BibitemOpen
  \bibfield  {author} {\bibinfo {author} {\bibfnamefont {N.~M.}\ \bibnamefont {Linke}}, \bibinfo {author} {\bibfnamefont {S.}~\bibnamefont {Johri}}, \bibinfo {author} {\bibfnamefont {C.}~\bibnamefont {Figgatt}}, \bibinfo {author} {\bibfnamefont {K.~A.}\ \bibnamefont {Landsman}}, \bibinfo {author} {\bibfnamefont {A.~Y.}\ \bibnamefont {Matsuura}}, \ and\ \bibinfo {author} {\bibfnamefont {C.}~\bibnamefont {Monroe}},\ }\href {\doibase 10.1103/physreva.98.052334} {\bibfield  {journal} {\bibinfo  {journal} {Physical Review A}\ }\textbf {\bibinfo {volume} {98}} (\bibinfo {year} {2018}),\ 10.1103/physreva.98.052334}\BibitemShut {NoStop}%
\bibitem [{\citenamefont {Horodecki}\ and\ \citenamefont {Ekert}(2002)}]{horodecki2002method}%
  \BibitemOpen
  \bibfield  {author} {\bibinfo {author} {\bibfnamefont {P.}~\bibnamefont {Horodecki}}\ and\ \bibinfo {author} {\bibfnamefont {A.}~\bibnamefont {Ekert}},\ }\href {\doibase 10.1103/physrevlett.89.127902} {\bibfield  {journal} {\bibinfo  {journal} {Physical Review Letters}\ }\textbf {\bibinfo {volume} {89}} (\bibinfo {year} {2002}),\ 10.1103/physrevlett.89.127902}\BibitemShut {NoStop}%
\bibitem [{\citenamefont {Ekert}\ \emph {et~al.}(2002)\citenamefont {Ekert}, \citenamefont {Alves}, \citenamefont {Oi}, \citenamefont {Horodecki}, \citenamefont {Horodecki},\ and\ \citenamefont {Kwek}}]{ekert2002direct}%
  \BibitemOpen
  \bibfield  {author} {\bibinfo {author} {\bibfnamefont {A.~K.}\ \bibnamefont {Ekert}}, \bibinfo {author} {\bibfnamefont {C.~M.}\ \bibnamefont {Alves}}, \bibinfo {author} {\bibfnamefont {D.~K.~L.}\ \bibnamefont {Oi}}, \bibinfo {author} {\bibfnamefont {M.}~\bibnamefont {Horodecki}}, \bibinfo {author} {\bibfnamefont {P.}~\bibnamefont {Horodecki}}, \ and\ \bibinfo {author} {\bibfnamefont {L.~C.}\ \bibnamefont {Kwek}},\ }\href {\doibase 10.1103/physrevlett.88.217901} {\bibfield  {journal} {\bibinfo  {journal} {Physical Review Letters}\ }\textbf {\bibinfo {volume} {88}} (\bibinfo {year} {2002}),\ 10.1103/physrevlett.88.217901}\BibitemShut {NoStop}%
\bibitem [{\citenamefont {Haug}\ \emph {et~al.}(2024{\natexlab{b}})\citenamefont {Haug}, \citenamefont {Bharti},\ and\ \citenamefont {Koh}}]{haug2024pseudorandomunitariesrealsparse}%
  \BibitemOpen
  \bibfield  {author} {\bibinfo {author} {\bibfnamefont {T.}~\bibnamefont {Haug}}, \bibinfo {author} {\bibfnamefont {K.}~\bibnamefont {Bharti}}, \ and\ \bibinfo {author} {\bibfnamefont {D.~E.}\ \bibnamefont {Koh}},\ }\href {https://arxiv.org/abs/2306.11677} {\bibfield  {journal} {\bibinfo  {journal} {arxiv:2306.11677}\ } (\bibinfo {year} {2024}{\natexlab{b}})}\BibitemShut {NoStop}%
\bibitem [{\citenamefont {Sandvik}(2003)}]{sandvik2003sse_ising}%
  \BibitemOpen
  \bibfield  {author} {\bibinfo {author} {\bibfnamefont {A.~W.}\ \bibnamefont {Sandvik}},\ }\href {\doibase 10.1103/physreve.68.056701} {\bibfield  {journal} {\bibinfo  {journal} {Physical Review E}\ }\textbf {\bibinfo {volume} {68}} (\bibinfo {year} {2003}),\ 10.1103/physreve.68.056701}\BibitemShut {NoStop}%
\bibitem [{sup()}]{supmat}%
  \BibitemOpen
  \href@noop {} {\bibinfo  {journal} {See Supplemental Material for technical details of the algorithm and additional numerical data, which includes Refs.~\cite{sandvik1999sse_loop,syljusen2002directedloops,syljusen2003directed,zhao2021higherform,sandvik1997finitesize,melko2005stochastic}}\ }\BibitemShut {NoStop}%
\bibitem [{\citenamefont {Sandvik}(2019)}]{sandvik2019sse}%
  \BibitemOpen
\bibfield  {journal} {  }\bibfield  {author} {\bibinfo {author} {\bibfnamefont {A.~W.}\ \bibnamefont {Sandvik}},\ }\href {https://arxiv.org/abs/1909.10591} {\bibfield  {journal} {\bibinfo  {journal} {arxiv:1909.10591}\ } (\bibinfo {year} {2019})}\BibitemShut {NoStop}%
\bibitem [{\citenamefont {Melko}(2013)}]{Melko2013sse}%
  \BibitemOpen
  \bibfield  {author} {\bibinfo {author} {\bibfnamefont {R.~G.}\ \bibnamefont {Melko}},\ }\enquote {\bibinfo {title} {Stochastic series expansion quantum monte carlo},}\ in\ \href {\doibase 10.1007/978-3-642-35106-8_7} {\emph {\bibinfo {booktitle} {Strongly Correlated Systems: Numerical Methods}}},\ \bibinfo {editor} {edited by\ \bibinfo {editor} {\bibfnamefont {A.}~\bibnamefont {Avella}}\ and\ \bibinfo {editor} {\bibfnamefont {F.}~\bibnamefont {Mancini}}}\ (\bibinfo  {publisher} {Springer Berlin Heidelberg},\ \bibinfo {address} {Berlin, Heidelberg},\ \bibinfo {year} {2013})\ pp.\ \bibinfo {pages} {185--206}\BibitemShut {NoStop}%
\bibitem [{\citenamefont {Di~Francesco}\ \emph {et~al.}(1997)\citenamefont {Di~Francesco}, \citenamefont {Mathieu},\ and\ \citenamefont {Sénéchal}}]{DiFrancesco}%
  \BibitemOpen
  \bibfield  {author} {\bibinfo {author} {\bibfnamefont {P.}~\bibnamefont {Di~Francesco}}, \bibinfo {author} {\bibfnamefont {P.}~\bibnamefont {Mathieu}}, \ and\ \bibinfo {author} {\bibfnamefont {D.}~\bibnamefont {Sénéchal}},\ }\href {\doibase 10.1007/978-1-4612-2256-9} {\emph {\bibinfo {title} {{Conformal field theory}}}},\ Graduate texts in contemporary physics\ (\bibinfo  {publisher} {Springer},\ \bibinfo {address} {New York, NY},\ \bibinfo {year} {1997})\BibitemShut {NoStop}%
\bibitem [{\citenamefont {Zeng}\ and\ \citenamefont {Wen}(2015)}]{zeng2015}%
  \BibitemOpen
  \bibfield  {author} {\bibinfo {author} {\bibfnamefont {B.}~\bibnamefont {Zeng}}\ and\ \bibinfo {author} {\bibfnamefont {X.-G.}\ \bibnamefont {Wen}},\ }\href {\doibase 10.1103/PhysRevB.91.125121} {\bibfield  {journal} {\bibinfo  {journal} {Phys. Rev. B}\ }\textbf {\bibinfo {volume} {91}},\ \bibinfo {pages} {125121} (\bibinfo {year} {2015})}\BibitemShut {NoStop}%
\bibitem [{\citenamefont {Zeng}\ \emph {et~al.}(2019)\citenamefont {Zeng}, \citenamefont {Chen}, \citenamefont {Zhou},\ and\ \citenamefont {Wen}}]{zeng2019}%
  \BibitemOpen
  \bibfield  {author} {\bibinfo {author} {\bibfnamefont {B.}~\bibnamefont {Zeng}}, \bibinfo {author} {\bibfnamefont {X.}~\bibnamefont {Chen}}, \bibinfo {author} {\bibfnamefont {D.-L.}\ \bibnamefont {Zhou}}, \ and\ \bibinfo {author} {\bibfnamefont {X.-G.}\ \bibnamefont {Wen}},\ }\href {\doibase 10.1007/978-1-4939-9084-9} {\emph {\bibinfo {title} {Quantum Information Meets Quantum Matter: From Quantum Entanglement to Topological Phases of Many-Body Systems}}}\ (\bibinfo  {publisher} {Springer New York},\ \bibinfo {year} {2019})\BibitemShut {NoStop}%
\bibitem [{\citenamefont {Lavasani}\ \emph {et~al.}(2021)\citenamefont {Lavasani}, \citenamefont {Alavirad},\ and\ \citenamefont {Barkeshli}}]{Lavasani2021}%
  \BibitemOpen
  \bibfield  {author} {\bibinfo {author} {\bibfnamefont {A.}~\bibnamefont {Lavasani}}, \bibinfo {author} {\bibfnamefont {Y.}~\bibnamefont {Alavirad}}, \ and\ \bibinfo {author} {\bibfnamefont {M.}~\bibnamefont {Barkeshli}},\ }\href {\doibase 10.1038/s41567-020-01112-z} {\bibfield  {journal} {\bibinfo  {journal} {Nature Physics}\ }\textbf {\bibinfo {volume} {17}},\ \bibinfo {pages} {342–347} (\bibinfo {year} {2021})}\BibitemShut {NoStop}%
\bibitem [{\citenamefont {Klocke}\ and\ \citenamefont {Buchhold}(2022)}]{Klocke2022}%
  \BibitemOpen
  \bibfield  {author} {\bibinfo {author} {\bibfnamefont {K.}~\bibnamefont {Klocke}}\ and\ \bibinfo {author} {\bibfnamefont {M.}~\bibnamefont {Buchhold}},\ }\href {\doibase 10.1103/physrevb.106.104307} {\bibfield  {journal} {\bibinfo  {journal} {Physical Review B}\ }\textbf {\bibinfo {volume} {106}} (\bibinfo {year} {2022}),\ 10.1103/physrevb.106.104307}\BibitemShut {NoStop}%
\bibitem [{\citenamefont {Bl\"{o}te}\ and\ \citenamefont {Deng}(2002)}]{blote2002cluster}%
  \BibitemOpen
  \bibfield  {author} {\bibinfo {author} {\bibfnamefont {H.~W.~J.}\ \bibnamefont {Bl\"{o}te}}\ and\ \bibinfo {author} {\bibfnamefont {Y.}~\bibnamefont {Deng}},\ }\href {\doibase 10.1103/physreve.66.066110} {\bibfield  {journal} {\bibinfo  {journal} {Physical Review E}\ }\textbf {\bibinfo {volume} {66}} (\bibinfo {year} {2002}),\ 10.1103/physreve.66.066110}\BibitemShut {NoStop}%
\bibitem [{\citenamefont {Zhao}\ \emph {et~al.}(2022)\citenamefont {Zhao}, \citenamefont {Chen}, \citenamefont {Wang}, \citenamefont {Yan}, \citenamefont {Cheng},\ and\ \citenamefont {Meng}}]{zhao2022measuring}%
  \BibitemOpen
  \bibfield  {author} {\bibinfo {author} {\bibfnamefont {J.}~\bibnamefont {Zhao}}, \bibinfo {author} {\bibfnamefont {B.-B.}\ \bibnamefont {Chen}}, \bibinfo {author} {\bibfnamefont {Y.-C.}\ \bibnamefont {Wang}}, \bibinfo {author} {\bibfnamefont {Z.}~\bibnamefont {Yan}}, \bibinfo {author} {\bibfnamefont {M.}~\bibnamefont {Cheng}}, \ and\ \bibinfo {author} {\bibfnamefont {Z.~Y.}\ \bibnamefont {Meng}},\ }\href {\doibase 10.1038/s41535-022-00476-0} {\bibfield  {journal} {\bibinfo  {journal} {npj Quantum Materials}\ }\textbf {\bibinfo {volume} {7}} (\bibinfo {year} {2022}),\ 10.1038/s41535-022-00476-0}\BibitemShut {NoStop}%
\bibitem [{\citenamefont {Garratt}\ \emph {et~al.}(2023)\citenamefont {Garratt}, \citenamefont {Weinstein},\ and\ \citenamefont {Altman}}]{garratt2023measurements}%
  \BibitemOpen
  \bibfield  {author} {\bibinfo {author} {\bibfnamefont {S.~J.}\ \bibnamefont {Garratt}}, \bibinfo {author} {\bibfnamefont {Z.}~\bibnamefont {Weinstein}}, \ and\ \bibinfo {author} {\bibfnamefont {E.}~\bibnamefont {Altman}},\ }\href {\doibase 10.1103/physrevx.13.021026} {\bibfield  {journal} {\bibinfo  {journal} {Physical Review X}\ }\textbf {\bibinfo {volume} {13}} (\bibinfo {year} {2023}),\ 10.1103/physrevx.13.021026}\BibitemShut {NoStop}%
\bibitem [{\citenamefont {Lee}\ \emph {et~al.}(2023)\citenamefont {Lee}, \citenamefont {Jian},\ and\ \citenamefont {Xu}}]{lee2023quantum}%
  \BibitemOpen
  \bibfield  {author} {\bibinfo {author} {\bibfnamefont {J.~Y.}\ \bibnamefont {Lee}}, \bibinfo {author} {\bibfnamefont {C.-M.}\ \bibnamefont {Jian}}, \ and\ \bibinfo {author} {\bibfnamefont {C.}~\bibnamefont {Xu}},\ }\href {\doibase 10.1103/prxquantum.4.030317} {\bibfield  {journal} {\bibinfo  {journal} {PRX Quantum}\ }\textbf {\bibinfo {volume} {4}} (\bibinfo {year} {2023}),\ 10.1103/prxquantum.4.030317}\BibitemShut {NoStop}%
\bibitem [{\citenamefont {Murciano}\ \emph {et~al.}(2023)\citenamefont {Murciano}, \citenamefont {Sala}, \citenamefont {Liu}, \citenamefont {Mong},\ and\ \citenamefont {Alicea}}]{murciano2023measurement}%
  \BibitemOpen
  \bibfield  {author} {\bibinfo {author} {\bibfnamefont {S.}~\bibnamefont {Murciano}}, \bibinfo {author} {\bibfnamefont {P.}~\bibnamefont {Sala}}, \bibinfo {author} {\bibfnamefont {Y.}~\bibnamefont {Liu}}, \bibinfo {author} {\bibfnamefont {R.~S.}\ \bibnamefont {Mong}}, \ and\ \bibinfo {author} {\bibfnamefont {J.}~\bibnamefont {Alicea}},\ }\href {\doibase 10.1103/physrevx.13.041042} {\bibfield  {journal} {\bibinfo  {journal} {Physical Review X}\ }\textbf {\bibinfo {volume} {13}} (\bibinfo {year} {2023}),\ 10.1103/physrevx.13.041042}\BibitemShut {NoStop}%
\bibitem [{\citenamefont {Weinstein}\ \emph {et~al.}(2023)\citenamefont {Weinstein}, \citenamefont {Sajith}, \citenamefont {Altman},\ and\ \citenamefont {Garratt}}]{weinstein2023nonlocality}%
  \BibitemOpen
  \bibfield  {author} {\bibinfo {author} {\bibfnamefont {Z.}~\bibnamefont {Weinstein}}, \bibinfo {author} {\bibfnamefont {R.}~\bibnamefont {Sajith}}, \bibinfo {author} {\bibfnamefont {E.}~\bibnamefont {Altman}}, \ and\ \bibinfo {author} {\bibfnamefont {S.~J.}\ \bibnamefont {Garratt}},\ }\href {\doibase 10.1103/physrevb.107.245132} {\bibfield  {journal} {\bibinfo  {journal} {Physical Review B}\ }\textbf {\bibinfo {volume} {107}} (\bibinfo {year} {2023}),\ 10.1103/physrevb.107.245132}\BibitemShut {NoStop}%
\bibitem [{\citenamefont {Yang}\ \emph {et~al.}(2023)\citenamefont {Yang}, \citenamefont {Mao},\ and\ \citenamefont {Jian}}]{yang2023entanglemet}%
  \BibitemOpen
  \bibfield  {author} {\bibinfo {author} {\bibfnamefont {Z.}~\bibnamefont {Yang}}, \bibinfo {author} {\bibfnamefont {D.}~\bibnamefont {Mao}}, \ and\ \bibinfo {author} {\bibfnamefont {C.-M.}\ \bibnamefont {Jian}},\ }\href {\doibase 10.1103/physrevb.108.165120} {\bibfield  {journal} {\bibinfo  {journal} {Physical Review B}\ }\textbf {\bibinfo {volume} {108}} (\bibinfo {year} {2023}),\ 10.1103/physrevb.108.165120}\BibitemShut {NoStop}%
\bibitem [{\citenamefont {Hoshino}\ \emph {et~al.}(2024)\citenamefont {Hoshino}, \citenamefont {Oshikawa},\ and\ \citenamefont {Ashida}}]{hoshino2024entanglementswapping}%
  \BibitemOpen
  \bibfield  {author} {\bibinfo {author} {\bibfnamefont {M.}~\bibnamefont {Hoshino}}, \bibinfo {author} {\bibfnamefont {M.}~\bibnamefont {Oshikawa}}, \ and\ \bibinfo {author} {\bibfnamefont {Y.}~\bibnamefont {Ashida}},\ }\href {https://arxiv.org/abs/2406.12377} {\bibfield  {journal} {\bibinfo  {journal} {arxiv:2406.12377}\ } (\bibinfo {year} {2024})}\BibitemShut {NoStop}%
\bibitem [{\citenamefont {Baweja}\ \emph {et~al.}(2024)\citenamefont {Baweja}, \citenamefont {Luitz},\ and\ \citenamefont {Garratt}}]{baweja2024postmeasurement}%
  \BibitemOpen
  \bibfield  {author} {\bibinfo {author} {\bibfnamefont {K.}~\bibnamefont {Baweja}}, \bibinfo {author} {\bibfnamefont {D.~J.}\ \bibnamefont {Luitz}}, \ and\ \bibinfo {author} {\bibfnamefont {S.~J.}\ \bibnamefont {Garratt}},\ }\href {https://arxiv.org/abs/2410.13844} {\bibfield  {journal} {\bibinfo  {journal} {arxiv:2410.13844}\ } (\bibinfo {year} {2024})}\BibitemShut {NoStop}%
\bibitem [{\citenamefont {Lessa}\ \emph {et~al.}(2025)\citenamefont {Lessa}, \citenamefont {Ma}, \citenamefont {Zhang}, \citenamefont {Bi}, \citenamefont {Cheng},\ and\ \citenamefont {Wang}}]{lessa2025strongtoweak}%
  \BibitemOpen
  \bibfield  {author} {\bibinfo {author} {\bibfnamefont {L.~A.}\ \bibnamefont {Lessa}}, \bibinfo {author} {\bibfnamefont {R.}~\bibnamefont {Ma}}, \bibinfo {author} {\bibfnamefont {J.-H.}\ \bibnamefont {Zhang}}, \bibinfo {author} {\bibfnamefont {Z.}~\bibnamefont {Bi}}, \bibinfo {author} {\bibfnamefont {M.}~\bibnamefont {Cheng}}, \ and\ \bibinfo {author} {\bibfnamefont {C.}~\bibnamefont {Wang}},\ }\href {\doibase 10.1103/prxquantum.6.010344} {\bibfield  {journal} {\bibinfo  {journal} {PRX Quantum}\ }\textbf {\bibinfo {volume} {6}} (\bibinfo {year} {2025}),\ 10.1103/prxquantum.6.010344}\BibitemShut {NoStop}%
\bibitem [{\citenamefont {Sala}\ \emph {et~al.}(2024)\citenamefont {Sala}, \citenamefont {Gopalakrishnan}, \citenamefont {Oshikawa},\ and\ \citenamefont {You}}]{sala2024spontaneous}%
  \BibitemOpen
  \bibfield  {author} {\bibinfo {author} {\bibfnamefont {P.}~\bibnamefont {Sala}}, \bibinfo {author} {\bibfnamefont {S.}~\bibnamefont {Gopalakrishnan}}, \bibinfo {author} {\bibfnamefont {M.}~\bibnamefont {Oshikawa}}, \ and\ \bibinfo {author} {\bibfnamefont {Y.}~\bibnamefont {You}},\ }\href {\doibase 10.1103/physrevb.110.155150} {\bibfield  {journal} {\bibinfo  {journal} {Physical Review B}\ }\textbf {\bibinfo {volume} {110}} (\bibinfo {year} {2024}),\ 10.1103/physrevb.110.155150}\BibitemShut {NoStop}%
\bibitem [{\citenamefont {Chitambar}\ and\ \citenamefont {Gour}(2019)}]{chitambar2019quantum}%
  \BibitemOpen
  \bibfield  {author} {\bibinfo {author} {\bibfnamefont {E.}~\bibnamefont {Chitambar}}\ and\ \bibinfo {author} {\bibfnamefont {G.}~\bibnamefont {Gour}},\ }\href {\doibase 10.1103/RevModPhys.91.025001} {\bibfield  {journal} {\bibinfo  {journal} {Rev. Mod. Phys.}\ }\textbf {\bibinfo {volume} {91}},\ \bibinfo {pages} {025001} (\bibinfo {year} {2019})}\BibitemShut {NoStop}%
\bibitem [{\citenamefont {Leone}\ \emph {et~al.}(2022)\citenamefont {Leone}, \citenamefont {Oliviero},\ and\ \citenamefont {Hamma}}]{leone2022stabilizer}%
  \BibitemOpen
  \bibfield  {author} {\bibinfo {author} {\bibfnamefont {L.}~\bibnamefont {Leone}}, \bibinfo {author} {\bibfnamefont {S.~F.~E.}\ \bibnamefont {Oliviero}}, \ and\ \bibinfo {author} {\bibfnamefont {A.}~\bibnamefont {Hamma}},\ }\href {\doibase 10.1103/PhysRevLett.128.050402} {\bibfield  {journal} {\bibinfo  {journal} {Phys. Rev. Lett.}\ }\textbf {\bibinfo {volume} {128}},\ \bibinfo {pages} {050402} (\bibinfo {year} {2022})}\BibitemShut {NoStop}%
\bibitem [{\citenamefont {Haug}\ and\ \citenamefont {Piroli}(2023{\natexlab{a}})}]{haug2022quantifying}%
  \BibitemOpen
  \bibfield  {author} {\bibinfo {author} {\bibfnamefont {T.}~\bibnamefont {Haug}}\ and\ \bibinfo {author} {\bibfnamefont {L.}~\bibnamefont {Piroli}},\ }\href {\doibase 10.1103/PhysRevB.107.035148} {\bibfield  {journal} {\bibinfo  {journal} {Phys. Rev. B}\ }\textbf {\bibinfo {volume} {107}},\ \bibinfo {pages} {035148} (\bibinfo {year} {2023}{\natexlab{a}})}\BibitemShut {NoStop}%
\bibitem [{\citenamefont {Lami}\ and\ \citenamefont {Collura}(2023)}]{lami2023nonstabilizerness}%
  \BibitemOpen
  \bibfield  {author} {\bibinfo {author} {\bibfnamefont {G.}~\bibnamefont {Lami}}\ and\ \bibinfo {author} {\bibfnamefont {M.}~\bibnamefont {Collura}},\ }\href {\doibase 10.1103/PhysRevLett.131.180401} {\bibfield  {journal} {\bibinfo  {journal} {Phys. Rev. Lett.}\ }\textbf {\bibinfo {volume} {131}},\ \bibinfo {pages} {180401} (\bibinfo {year} {2023})}\BibitemShut {NoStop}%
\bibitem [{\citenamefont {Haug}\ and\ \citenamefont {Piroli}(2023{\natexlab{b}})}]{haug2023stabilizer}%
  \BibitemOpen
  \bibfield  {author} {\bibinfo {author} {\bibfnamefont {T.}~\bibnamefont {Haug}}\ and\ \bibinfo {author} {\bibfnamefont {L.}~\bibnamefont {Piroli}},\ }\href {\doibase 10.22331/q-2023-08-28-1092} {\bibfield  {journal} {\bibinfo  {journal} {Quantum}\ }\textbf {\bibinfo {volume} {7}},\ \bibinfo {pages} {1092} (\bibinfo {year} {2023}{\natexlab{b}})}\BibitemShut {NoStop}%
\bibitem [{\citenamefont {Tarabunga}\ \emph {et~al.}(2023)\citenamefont {Tarabunga}, \citenamefont {Tirrito}, \citenamefont {Chanda},\ and\ \citenamefont {Dalmonte}}]{tarabunga2023many}%
  \BibitemOpen
  \bibfield  {author} {\bibinfo {author} {\bibfnamefont {P.~S.}\ \bibnamefont {Tarabunga}}, \bibinfo {author} {\bibfnamefont {E.}~\bibnamefont {Tirrito}}, \bibinfo {author} {\bibfnamefont {T.}~\bibnamefont {Chanda}}, \ and\ \bibinfo {author} {\bibfnamefont {M.}~\bibnamefont {Dalmonte}},\ }\href {\doibase 10.1103/PRXQuantum.4.040317} {\bibfield  {journal} {\bibinfo  {journal} {PRX Quantum}\ }\textbf {\bibinfo {volume} {4}},\ \bibinfo {pages} {040317} (\bibinfo {year} {2023})}\BibitemShut {NoStop}%
\bibitem [{\citenamefont {Tarabunga}\ \emph {et~al.}(2024)\citenamefont {Tarabunga}, \citenamefont {Tirrito}, \citenamefont {Bañuls},\ and\ \citenamefont {Dalmonte}}]{tarabunga2024nonstabilizernessmps}%
  \BibitemOpen
  \bibfield  {author} {\bibinfo {author} {\bibfnamefont {P.~S.}\ \bibnamefont {Tarabunga}}, \bibinfo {author} {\bibfnamefont {E.}~\bibnamefont {Tirrito}}, \bibinfo {author} {\bibfnamefont {M.~C.}\ \bibnamefont {Bañuls}}, \ and\ \bibinfo {author} {\bibfnamefont {M.}~\bibnamefont {Dalmonte}},\ }\href {\doibase 10.1103/PhysRevLett.133.010601} {\bibfield  {journal} {\bibinfo  {journal} {Phys. Rev. Lett.}\ }\textbf {\bibinfo {volume} {133}},\ \bibinfo {pages} {010601} (\bibinfo {year} {2024})}\BibitemShut {NoStop}%
\bibitem [{\citenamefont {Tarabunga}(2024)}]{tarabunga2024critical}%
  \BibitemOpen
  \bibfield  {author} {\bibinfo {author} {\bibfnamefont {P.~S.}\ \bibnamefont {Tarabunga}},\ }\href {https://doi.org/10.22331/q-2024-07-17-1413} {\bibfield  {journal} {\bibinfo  {journal} {Quantum}\ }\textbf {\bibinfo {volume} {8}},\ \bibinfo {pages} {1413} (\bibinfo {year} {2024})}\BibitemShut {NoStop}%
\bibitem [{\citenamefont {Tarabunga}\ and\ \citenamefont {Castelnovo}(2024)}]{tarabunga2023magic}%
  \BibitemOpen
  \bibfield  {author} {\bibinfo {author} {\bibfnamefont {P.~S.}\ \bibnamefont {Tarabunga}}\ and\ \bibinfo {author} {\bibfnamefont {C.}~\bibnamefont {Castelnovo}},\ }\href {\doibase 10.22331/q-2024-05-14-1347} {\bibfield  {journal} {\bibinfo  {journal} {Quantum}\ }\textbf {\bibinfo {volume} {8}},\ \bibinfo {pages} {1347} (\bibinfo {year} {2024})}\BibitemShut {NoStop}%
\bibitem [{\citenamefont {Falcão}\ \emph {et~al.}(2025)\citenamefont {Falcão}, \citenamefont {Tarabunga}, \citenamefont {Frau}, \citenamefont {Tirrito}, \citenamefont {Zakrzewski},\ and\ \citenamefont {Dalmonte}}]{falcao2024nonstabilizerness}%
  \BibitemOpen
  \bibfield  {author} {\bibinfo {author} {\bibfnamefont {P.~R.~N.}\ \bibnamefont {Falcão}}, \bibinfo {author} {\bibfnamefont {P.~S.}\ \bibnamefont {Tarabunga}}, \bibinfo {author} {\bibfnamefont {M.}~\bibnamefont {Frau}}, \bibinfo {author} {\bibfnamefont {E.}~\bibnamefont {Tirrito}}, \bibinfo {author} {\bibfnamefont {J.}~\bibnamefont {Zakrzewski}}, \ and\ \bibinfo {author} {\bibfnamefont {M.}~\bibnamefont {Dalmonte}},\ }\href {\doibase 10.1103/physrevb.111.l081102} {\bibfield  {journal} {\bibinfo  {journal} {Physical Review B}\ }\textbf {\bibinfo {volume} {111}} (\bibinfo {year} {2025}),\ 10.1103/physrevb.111.l081102}\BibitemShut {NoStop}%
\bibitem [{\citenamefont {Tarabunga}\ and\ \citenamefont {Haug}(2025{\natexlab{a}})}]{tarabunga2025efficient}%
  \BibitemOpen
  \bibfield  {author} {\bibinfo {author} {\bibfnamefont {P.~S.}\ \bibnamefont {Tarabunga}}\ and\ \bibinfo {author} {\bibfnamefont {T.}~\bibnamefont {Haug}},\ }\href {https://arxiv.org/abs/2504.07230} {\bibfield  {journal} {\bibinfo  {journal} {arXiv:2504.07230}\ } (\bibinfo {year} {2025}{\natexlab{a}})}\BibitemShut {NoStop}%
\bibitem [{\citenamefont {Collura}\ \emph {et~al.}(2024)\citenamefont {Collura}, \citenamefont {Nardis}, \citenamefont {Alba},\ and\ \citenamefont {Lami}}]{collura2024quantum}%
  \BibitemOpen
  \bibfield  {author} {\bibinfo {author} {\bibfnamefont {M.}~\bibnamefont {Collura}}, \bibinfo {author} {\bibfnamefont {J.~D.}\ \bibnamefont {Nardis}}, \bibinfo {author} {\bibfnamefont {V.}~\bibnamefont {Alba}}, \ and\ \bibinfo {author} {\bibfnamefont {G.}~\bibnamefont {Lami}},\ }\href {https://arxiv.org/abs/2412.05367} {\bibfield  {journal} {\bibinfo  {journal} {arxiv:2412.05367}\ } (\bibinfo {year} {2024})}\BibitemShut {NoStop}%
\bibitem [{\citenamefont {Liu}\ and\ \citenamefont {Clark}(2025)}]{liu2025nonequilibrium}%
  \BibitemOpen
  \bibfield  {author} {\bibinfo {author} {\bibfnamefont {Z.}~\bibnamefont {Liu}}\ and\ \bibinfo {author} {\bibfnamefont {B.~K.}\ \bibnamefont {Clark}},\ }\href {\doibase 10.1103/physrevb.111.085144} {\bibfield  {journal} {\bibinfo  {journal} {Physical Review B}\ }\textbf {\bibinfo {volume} {111}} (\bibinfo {year} {2025}),\ 10.1103/physrevb.111.085144}\BibitemShut {NoStop}%
\bibitem [{\citenamefont {Ding}\ \emph {et~al.}(2025)\citenamefont {Ding}, \citenamefont {Wang},\ and\ \citenamefont {Yan}}]{ding2025evaluating}%
  \BibitemOpen
  \bibfield  {author} {\bibinfo {author} {\bibfnamefont {Y.-M.}\ \bibnamefont {Ding}}, \bibinfo {author} {\bibfnamefont {Z.}~\bibnamefont {Wang}}, \ and\ \bibinfo {author} {\bibfnamefont {Z.}~\bibnamefont {Yan}},\ }\href {\doibase 10.1103/pyzr-jmvw} {\bibfield  {journal} {\bibinfo  {journal} {PRX Quantum}\ }\textbf {\bibinfo {volume} {6}},\ \bibinfo {pages} {030328} (\bibinfo {year} {2025})}\BibitemShut {NoStop}%
\bibitem [{\citenamefont {Sandvik}(1998)}]{Sandvik1998sac}%
  \BibitemOpen
  \bibfield  {author} {\bibinfo {author} {\bibfnamefont {A.~W.}\ \bibnamefont {Sandvik}},\ }\href {\doibase 10.1103/PhysRevB.57.10287} {\bibfield  {journal} {\bibinfo  {journal} {Phys. Rev. B}\ }\textbf {\bibinfo {volume} {57}},\ \bibinfo {pages} {10287} (\bibinfo {year} {1998})}\BibitemShut {NoStop}%
\bibitem [{\citenamefont {Sylju\aa{}sen}(2008)}]{Syljuasen2008sac}%
  \BibitemOpen
  \bibfield  {author} {\bibinfo {author} {\bibfnamefont {O.~F.}\ \bibnamefont {Sylju\aa{}sen}},\ }\href {\doibase 10.1103/PhysRevB.78.174429} {\bibfield  {journal} {\bibinfo  {journal} {Phys. Rev. B}\ }\textbf {\bibinfo {volume} {78}},\ \bibinfo {pages} {174429} (\bibinfo {year} {2008})}\BibitemShut {NoStop}%
\bibitem [{\citenamefont {Sandvik}(2016)}]{Sandvik2016sac}%
  \BibitemOpen
  \bibfield  {author} {\bibinfo {author} {\bibfnamefont {A.~W.}\ \bibnamefont {Sandvik}},\ }\href {\doibase 10.1103/PhysRevE.94.063308} {\bibfield  {journal} {\bibinfo  {journal} {Phys. Rev. E}\ }\textbf {\bibinfo {volume} {94}},\ \bibinfo {pages} {063308} (\bibinfo {year} {2016})}\BibitemShut {NoStop}%
\bibitem [{\citenamefont {Shao}\ and\ \citenamefont {Sandvik}(2023)}]{Shao2023sac}%
  \BibitemOpen
  \bibfield  {author} {\bibinfo {author} {\bibfnamefont {H.}~\bibnamefont {Shao}}\ and\ \bibinfo {author} {\bibfnamefont {A.~W.}\ \bibnamefont {Sandvik}},\ }\href {\doibase https://doi.org/10.1016/j.physrep.2022.11.002} {\bibfield  {journal} {\bibinfo  {journal} {Physics Reports}\ }\textbf {\bibinfo {volume} {1003}},\ \bibinfo {pages} {1} (\bibinfo {year} {2023})},\ \bibinfo {note} {progress on stochastic analytic continuation of quantum Monte Carlo data}\BibitemShut {NoStop}%
\bibitem [{\citenamefont {Calabrese}\ \emph {et~al.}(2012)\citenamefont {Calabrese}, \citenamefont {Cardy},\ and\ \citenamefont {Tonni}}]{calabrese2012entanglement}%
  \BibitemOpen
  \bibfield  {author} {\bibinfo {author} {\bibfnamefont {P.}~\bibnamefont {Calabrese}}, \bibinfo {author} {\bibfnamefont {J.}~\bibnamefont {Cardy}}, \ and\ \bibinfo {author} {\bibfnamefont {E.}~\bibnamefont {Tonni}},\ }\href {\doibase 10.1103/physrevlett.109.130502} {\bibfield  {journal} {\bibinfo  {journal} {Physical Review Letters}\ }\textbf {\bibinfo {volume} {109}} (\bibinfo {year} {2012}),\ 10.1103/physrevlett.109.130502}\BibitemShut {NoStop}%
\bibitem [{\citenamefont {Ding}\ \emph {et~al.}(2024{\natexlab{b}})\citenamefont {Ding}, \citenamefont {Tang}, \citenamefont {Wang}, \citenamefont {Wang}, \citenamefont {Mao},\ and\ \citenamefont {Yan}}]{ding2024negativity}%
  \BibitemOpen
  \bibfield  {author} {\bibinfo {author} {\bibfnamefont {Y.-M.}\ \bibnamefont {Ding}}, \bibinfo {author} {\bibfnamefont {Y.}~\bibnamefont {Tang}}, \bibinfo {author} {\bibfnamefont {Z.}~\bibnamefont {Wang}}, \bibinfo {author} {\bibfnamefont {Z.}~\bibnamefont {Wang}}, \bibinfo {author} {\bibfnamefont {B.-B.}\ \bibnamefont {Mao}}, \ and\ \bibinfo {author} {\bibfnamefont {Z.}~\bibnamefont {Yan}},\ }\href {https://arxiv.org/abs/2409.10273} {\bibfield  {journal} {\bibinfo  {journal} {arXiv:2409.10273}\ } (\bibinfo {year} {2024}{\natexlab{b}})}\BibitemShut {NoStop}%
\bibitem [{\citenamefont {Wu}\ \emph {et~al.}(2020)\citenamefont {Wu}, \citenamefont {Lu}, \citenamefont {Chung}, \citenamefont {Kao},\ and\ \citenamefont {Grover}}]{wukaihsin2020negativity}%
  \BibitemOpen
  \bibfield  {author} {\bibinfo {author} {\bibfnamefont {K.-H.}\ \bibnamefont {Wu}}, \bibinfo {author} {\bibfnamefont {T.-C.}\ \bibnamefont {Lu}}, \bibinfo {author} {\bibfnamefont {C.-M.}\ \bibnamefont {Chung}}, \bibinfo {author} {\bibfnamefont {Y.-J.}\ \bibnamefont {Kao}}, \ and\ \bibinfo {author} {\bibfnamefont {T.}~\bibnamefont {Grover}},\ }\href {\doibase 10.1103/PhysRevLett.125.140603} {\bibfield  {journal} {\bibinfo  {journal} {Phys. Rev. Lett.}\ }\textbf {\bibinfo {volume} {125}},\ \bibinfo {pages} {140603} (\bibinfo {year} {2020})}\BibitemShut {NoStop}%
\bibitem [{\citenamefont {Wang}\ and\ \citenamefont {Xu}(2025)}]{wangfohong2025negativity}%
  \BibitemOpen
  \bibfield  {author} {\bibinfo {author} {\bibfnamefont {F.-H.}\ \bibnamefont {Wang}}\ and\ \bibinfo {author} {\bibfnamefont {X.~Y.}\ \bibnamefont {Xu}},\ }\href {\doibase 10.1038/s41467-025-57971-8} {\bibfield  {journal} {\bibinfo  {journal} {Nature Communications}\ }\textbf {\bibinfo {volume} {16}},\ \bibinfo {pages} {2637} (\bibinfo {year} {2025})}\BibitemShut {NoStop}%
\bibitem [{\citenamefont {Chung}\ \emph {et~al.}(2014)\citenamefont {Chung}, \citenamefont {Alba}, \citenamefont {Bonnes}, \citenamefont {Chen},\ and\ \citenamefont {L\"{a}uchli}}]{chung2014entanglement}%
  \BibitemOpen
  \bibfield  {author} {\bibinfo {author} {\bibfnamefont {C.-M.}\ \bibnamefont {Chung}}, \bibinfo {author} {\bibfnamefont {V.}~\bibnamefont {Alba}}, \bibinfo {author} {\bibfnamefont {L.}~\bibnamefont {Bonnes}}, \bibinfo {author} {\bibfnamefont {P.}~\bibnamefont {Chen}}, \ and\ \bibinfo {author} {\bibfnamefont {A.~M.}\ \bibnamefont {L\"{a}uchli}},\ }\href {\doibase 10.1103/physrevb.90.064401} {\bibfield  {journal} {\bibinfo  {journal} {Physical Review B}\ }\textbf {\bibinfo {volume} {90}} (\bibinfo {year} {2014}),\ 10.1103/physrevb.90.064401}\BibitemShut {NoStop}%
\bibitem [{\citenamefont {Elben}\ \emph {et~al.}(2020)\citenamefont {Elben}, \citenamefont {Kueng}, \citenamefont {Huang}, \citenamefont {van Bijnen}, \citenamefont {Kokail}, \citenamefont {Dalmonte}, \citenamefont {Calabrese}, \citenamefont {Kraus}, \citenamefont {Preskill}, \citenamefont {Zoller},\ and\ \citenamefont {Vermersch}}]{Elben2020pkppt}%
  \BibitemOpen
  \bibfield  {author} {\bibinfo {author} {\bibfnamefont {A.}~\bibnamefont {Elben}}, \bibinfo {author} {\bibfnamefont {R.}~\bibnamefont {Kueng}}, \bibinfo {author} {\bibfnamefont {H.-Y.~R.}\ \bibnamefont {Huang}}, \bibinfo {author} {\bibfnamefont {R.}~\bibnamefont {van Bijnen}}, \bibinfo {author} {\bibfnamefont {C.}~\bibnamefont {Kokail}}, \bibinfo {author} {\bibfnamefont {M.}~\bibnamefont {Dalmonte}}, \bibinfo {author} {\bibfnamefont {P.}~\bibnamefont {Calabrese}}, \bibinfo {author} {\bibfnamefont {B.}~\bibnamefont {Kraus}}, \bibinfo {author} {\bibfnamefont {J.}~\bibnamefont {Preskill}}, \bibinfo {author} {\bibfnamefont {P.}~\bibnamefont {Zoller}}, \ and\ \bibinfo {author} {\bibfnamefont {B.}~\bibnamefont {Vermersch}},\ }\href {\doibase 10.1103/PhysRevLett.125.200501} {\bibfield  {journal} {\bibinfo  {journal} {Phys. Rev. Lett.}\ }\textbf {\bibinfo {volume} {125}},\ \bibinfo {pages} {200501} (\bibinfo {year} {2020})}\BibitemShut {NoStop}%
\bibitem [{\citenamefont {Neven}\ \emph {et~al.}(2021)\citenamefont {Neven}, \citenamefont {Carrasco}, \citenamefont {Vitale}, \citenamefont {Kokail}, \citenamefont {Elben}, \citenamefont {Dalmonte}, \citenamefont {Calabrese}, \citenamefont {Zoller}, \citenamefont {Vermersch}, \citenamefont {Kueng},\ and\ \citenamefont {Kraus}}]{Neven2021pkppt}%
  \BibitemOpen
  \bibfield  {author} {\bibinfo {author} {\bibfnamefont {A.}~\bibnamefont {Neven}}, \bibinfo {author} {\bibfnamefont {J.}~\bibnamefont {Carrasco}}, \bibinfo {author} {\bibfnamefont {V.}~\bibnamefont {Vitale}}, \bibinfo {author} {\bibfnamefont {C.}~\bibnamefont {Kokail}}, \bibinfo {author} {\bibfnamefont {A.}~\bibnamefont {Elben}}, \bibinfo {author} {\bibfnamefont {M.}~\bibnamefont {Dalmonte}}, \bibinfo {author} {\bibfnamefont {P.}~\bibnamefont {Calabrese}}, \bibinfo {author} {\bibfnamefont {P.}~\bibnamefont {Zoller}}, \bibinfo {author} {\bibfnamefont {B.}~\bibnamefont {Vermersch}}, \bibinfo {author} {\bibfnamefont {R.}~\bibnamefont {Kueng}}, \ and\ \bibinfo {author} {\bibfnamefont {B.}~\bibnamefont {Kraus}},\ }\href {\doibase 10.1038/s41534-021-00487-y} {\bibfield  {journal} {\bibinfo  {journal} {npj Quantum Information}\ }\textbf {\bibinfo {volume} {7}},\ \bibinfo {pages} {152} (\bibinfo {year} {2021})}\BibitemShut {NoStop}%
\bibitem [{\citenamefont {Yu}\ \emph {et~al.}(2021)\citenamefont {Yu}, \citenamefont {Imai},\ and\ \citenamefont {G\"{u}hne}}]{Yu2021}%
  \BibitemOpen
  \bibfield  {author} {\bibinfo {author} {\bibfnamefont {X.-D.}\ \bibnamefont {Yu}}, \bibinfo {author} {\bibfnamefont {S.}~\bibnamefont {Imai}}, \ and\ \bibinfo {author} {\bibfnamefont {O.}~\bibnamefont {G\"{u}hne}},\ }\href {\doibase 10.1103/physrevlett.127.060504} {\bibfield  {journal} {\bibinfo  {journal} {Physical Review Letters}\ }\textbf {\bibinfo {volume} {127}} (\bibinfo {year} {2021}),\ 10.1103/physrevlett.127.060504}\BibitemShut {NoStop}%
\bibitem [{\citenamefont {Tarabunga}\ and\ \citenamefont {Haug}(2025{\natexlab{b}})}]{tarabunga2025quantifying}%
  \BibitemOpen
  \bibfield  {author} {\bibinfo {author} {\bibfnamefont {P.~S.}\ \bibnamefont {Tarabunga}}\ and\ \bibinfo {author} {\bibfnamefont {T.}~\bibnamefont {Haug}},\ }\href {https://arxiv.org/abs/2507.13840} {\bibfield  {journal} {\bibinfo  {journal} {arxiv:2507.13840}\ } (\bibinfo {year} {2025}{\natexlab{b}})}\BibitemShut {NoStop}%
\bibitem [{\citenamefont {Tarabunga}\ and\ \citenamefont {Ding}(2025)}]{zenodo}%
  \BibitemOpen
  \bibfield  {author} {\bibinfo {author} {\bibfnamefont {P.~S.}\ \bibnamefont {Tarabunga}}\ and\ \bibinfo {author} {\bibfnamefont {Y.-M.}\ \bibnamefont {Ding}},\ }\href {\doibase 10.5281/zenodo.15470953} {\enquote {\bibinfo {title} {Bell sampling in quantum monte carlo simulations},}\ } (\bibinfo {year} {2025})\BibitemShut {NoStop}%
\bibitem [{\citenamefont {Sandvik}(1999)}]{sandvik1999sse_loop}%
  \BibitemOpen
  \bibfield  {author} {\bibinfo {author} {\bibfnamefont {A.~W.}\ \bibnamefont {Sandvik}},\ }\href {\doibase 10.1103/physrevb.59.r14157} {\bibfield  {journal} {\bibinfo  {journal} {Physical Review B}\ }\textbf {\bibinfo {volume} {59}},\ \bibinfo {pages} {R14157–R14160} (\bibinfo {year} {1999})}\BibitemShut {NoStop}%
\bibitem [{\citenamefont {Syljuåsen}\ and\ \citenamefont {Sandvik}(2002)}]{syljusen2002directedloops}%
  \BibitemOpen
  \bibfield  {author} {\bibinfo {author} {\bibfnamefont {O.~F.}\ \bibnamefont {Syljuåsen}}\ and\ \bibinfo {author} {\bibfnamefont {A.~W.}\ \bibnamefont {Sandvik}},\ }\href {\doibase 10.1103/physreve.66.046701} {\bibfield  {journal} {\bibinfo  {journal} {Physical Review E}\ }\textbf {\bibinfo {volume} {66}} (\bibinfo {year} {2002}),\ 10.1103/physreve.66.046701}\BibitemShut {NoStop}%
\bibitem [{\citenamefont {Syljuåsen}(2003)}]{syljusen2003directed}%
  \BibitemOpen
  \bibfield  {author} {\bibinfo {author} {\bibfnamefont {O.~F.}\ \bibnamefont {Syljuåsen}},\ }\href {\doibase 10.1103/physreve.67.046701} {\bibfield  {journal} {\bibinfo  {journal} {Physical Review E}\ }\textbf {\bibinfo {volume} {67}} (\bibinfo {year} {2003}),\ 10.1103/physreve.67.046701}\BibitemShut {NoStop}%
\bibitem [{\citenamefont {Zhao}\ \emph {et~al.}(2021)\citenamefont {Zhao}, \citenamefont {Yan}, \citenamefont {Cheng},\ and\ \citenamefont {Meng}}]{zhao2021higherform}%
  \BibitemOpen
  \bibfield  {author} {\bibinfo {author} {\bibfnamefont {J.}~\bibnamefont {Zhao}}, \bibinfo {author} {\bibfnamefont {Z.}~\bibnamefont {Yan}}, \bibinfo {author} {\bibfnamefont {M.}~\bibnamefont {Cheng}}, \ and\ \bibinfo {author} {\bibfnamefont {Z.~Y.}\ \bibnamefont {Meng}},\ }\href {\doibase 10.1103/physrevresearch.3.033024} {\bibfield  {journal} {\bibinfo  {journal} {Physical Review Research}\ }\textbf {\bibinfo {volume} {3}} (\bibinfo {year} {2021}),\ 10.1103/physrevresearch.3.033024}\BibitemShut {NoStop}%
\bibitem [{\citenamefont {Sandvik}(1997)}]{sandvik1997finitesize}%
  \BibitemOpen
  \bibfield  {author} {\bibinfo {author} {\bibfnamefont {A.~W.}\ \bibnamefont {Sandvik}},\ }\href {\doibase 10.1103/physrevb.56.11678} {\bibfield  {journal} {\bibinfo  {journal} {Physical Review B}\ }\textbf {\bibinfo {volume} {56}},\ \bibinfo {pages} {11678–11690} (\bibinfo {year} {1997})}\BibitemShut {NoStop}%
\bibitem [{\citenamefont {Melko}\ and\ \citenamefont {Sandvik}(2005)}]{melko2005stochastic}%
  \BibitemOpen
  \bibfield  {author} {\bibinfo {author} {\bibfnamefont {R.~G.}\ \bibnamefont {Melko}}\ and\ \bibinfo {author} {\bibfnamefont {A.~W.}\ \bibnamefont {Sandvik}},\ }\href {\doibase 10.1103/physreve.72.026702} {\bibfield  {journal} {\bibinfo  {journal} {Physical Review E}\ }\textbf {\bibinfo {volume} {72}} (\bibinfo {year} {2005}),\ 10.1103/physreve.72.026702}\BibitemShut {NoStop}%
\end{thebibliography}%

\section{End Matter}

\subsection{Thermodynamic integration} \label{sec:thermo}

We write the purity as
\begin{equation}
    \Tr\rho_A^2 = \frac{1}{2^{N_A}} \sum_{P \in \mathcal{P}_A}  \lvert \Tr(\rho P) \rvert^2,
\end{equation}
where $\mathcal{P}_A$ is the set of Pauli strings in $A$ (with identity elsewhere) and $N_A$ is the number of sites in $A$. Restricting to pure states, the Bell samples can be associated to Pauli strings with the distribution $P(\mathbf{r}) = \frac{1}{2^N} | \bra{\psi} \sigma_\mathbf{r} \ket{\psi}|^2$\footnote{More precisely, one needs to measure from $\ket{\psi} \otimes \ket{\psi^*}$ to sample the Pauli strings. If the Hamiltonian $H$ is not time-reversal symmetric, the Bell-basis Hamiltonian should be modified to $\HH=H\otimes I + I \otimes H^*$.}. Thus, $S_2$ can be expressed as
\begin{equation} \label{eq:estimator_alt}
    S_2(A) = -\ln  \frac{Q_A}{2^{N_A}  Q_{\varnothing}},
\end{equation}
where we define
\begin{equation}
    Q_A =  \sum_{\{r^z, r^x\}}  \bra{r^z_A,r^x_A, I_{\Bar{A}}} e^{- \beta \HH} \ket{r^z_A,r^x_A, I_{\Bar{A}}}
\end{equation}
for any subsystem $A$ and its complement $\Bar{A}$. Here, $I_{\Bar{A}}$ represents the Bell state $\ket{0,0}$ for all sites in $\Bar{A}$ and $\varnothing$ denotes the subsystem with no sites. Notice that, compared to the ordinary Bell basis partition function $\Tilde{Z}_B$, $Q_A$ can be simulated by freezing the sites in $\Bar{A}$ to $\ket{0,0}$. This can be achieved simply by not flipping any (bond) cluster which contains sites outside $A$.

In principle, Eq.~\eqref{eq:estimator_alt} presents an alternative way to estimate $S_2$, by counting the number of times that the sampled Bell states in $A$ simulated according to $Q_A$ are all $\ket{0,0}$. It is clear that the probability is exponentially small in $N_A$ regardless of the amount of the entanglement. Thus, the direct estimation would have significantly worse performance than Eq.~\eqref{eq:estimator_s2}. Nevertheless, $Q_A$ is now free from the sign problem, enabling us to apply specialized techniques to mitigate the error.

Following Ref.~\cite{emidio2020entanglement}, we introduce an extended ensemble 
\begin{equation} \label{eq:extended_ensemble}
    Q(\lambda) = \sum_{B \subseteq A} \lambda^{N_B} (1-\lambda)^{N_A - N_B} Q_B,
\end{equation}
 such that $Q(0)=Q_{\varnothing}$ and $Q(1) = Q_A$. With this, the R\'enyi-2 entanglement can be rewritten as
 \begin{equation}\label{eq:s2_thm_int}
    S_2 = -\ln \frac{Q(1)}{2^{N_A}Q(0)} = -\int_{0}^1d\lambda \frac{\partial \ln Q(\lambda)}{\partial\lambda}+N_A\ln 2,
\end{equation}
where the integrand is given by estimator $\langle e_1\rangle_{\lambda}$, which is defined as 
\begin{equation}
    \langle e_1\rangle_{\lambda} = \frac{\partial \ln Q(\lambda)}{\partial \lambda} =\bigg \langle \frac{N_B}{\lambda} -\frac{N_A-N_B}{1-\lambda}\bigg\rangle_{\lambda}.
\end{equation}
This integral can then be evaluated by numerical integration, analogous to the thermodynamic integration approach to calculate free energy in classical systems. Other methods such as non-equilibrium Jarzynski's equation~\cite{emidio2020entanglement} can also be used.

In Eq.~\eqref{eq:extended_ensemble}, sampling involves both the usual SSE configurations and the configurations of $B$. In Bell-QMC, the sum over $B \subseteq A$ can be performed analytically, resulting in an expression of $Q(\lambda)$ in terms of the Pauli strings as
\begin{equation}\label{eq:new_extended_ensemble}
    Q(\lambda) = \sum_{P \in \mathcal{P}_A} \lambda^{\text{wt}(P)} \lvert  \Tr(e^{-\beta H} P) \rvert^2,
\end{equation}
where $\text{wt}(P)$ is the number of non-identity Pauli operators, i.e., the weight, of the Pauli string $P$. Thus, the integrand can also be estimated with a different estimator $\langle e_2\rangle_{\lambda}$, defined by 
\begin{equation}
     \langle e_2\rangle_{\lambda} = \frac{\partial \ln Q(\lambda)}{\partial \lambda} =\frac{\langle {\text{wt}(P)}\rangle_{\lambda}}{{\lambda}}.
\end{equation}

The algorithms for simulating Eq.~\eqref{eq:extended_ensemble} and~\eqref{eq:new_extended_ensemble} are discussed in the Supplemental Material~\cite{supmat}. Notably, Eq.~\eqref{eq:new_extended_ensemble}  effectively reduces the sampling space back to the usual SSE configurations, albeit with modified weights. Consequently, we argue that it offers an advantage over Eq.~\eqref{eq:extended_ensemble} by simplifying the configuration space. We corroborate this argument by showing that $\langle e_2\rangle_{\lambda}$ has much smaller variance compared to $\langle e_1\rangle_{\lambda}$, as detailed in the Supplemental Material~\cite{supmat}. Note that this analytical summation over $B$ is generally not possible in the conventional QMC. This indicates that even in this scenario, Bell-QMC remains advantageous over computational-basis QMC.

We remark that, from Eq.~\eqref{eq:new_extended_ensemble}, one can see that the partition function $Q(\lambda)$ can be obtained from single-qubit depolarization noise channel $\Gamma_i(\rho)=\lambda\rho+(1-\lambda) \text{tr}_i(\rho) I_i/2$ with depolarization strength $1-\lambda$ applied to $e^{-\beta H}$ at all sites.

\newpage 
\onecolumngrid
\cleardoublepage
\setcounter{section}{0}
\setcounter{page}{1}
\setcounter{equation}{0}
\thispagestyle{empty}
\begin{center}
\textbf{\large Supplemental Material for ``Bell sampling in Quantum Monte Carlo simulations''}\\
\vspace{2ex}
Poetri Sonya Tarabunga and Yi-Ming Ding
\vspace{2ex}
\end{center}

\twocolumngrid
\pagenumbering{roman}

\setcounter{equation}{0}
\setcounter{figure}{0}
\renewcommand{\theequation}{S\arabic{equation}}
\renewcommand{\thefigure}{S\arabic{figure}}
\renewcommand{\thetable}{S\arabic{table}}

\section{Brief review of Bell sampling} \label{sec:bell_sampling_s,}
We define the Pauli matrices as $\sigma_{0 0} = I$, $\sigma_{01} = X$, $\sigma_{10} = Z$, and $\sigma_{11} =i Y$. The Bell states are defined as $\ket{\sigma_r} = (\sigma_r \otimes I ) \ket{\Phi^+}$ for $r=\{r^z,r^x\} \in \{0,1\}^2$, where $\ket{\Phi^+} = (\ket{00} + \ket{11})/\sqrt{2}$. Specifically,
\begin{align} \label{eq:bell_states}
    \ket{\sigma_{00}} = (\ket{00} + \ket{11})/\sqrt{2}, \\
    \ket{\sigma_{01}} = (\ket{10} + \ket{01})/\sqrt{2}, \\
    \ket{\sigma_{10}} = (\ket{00} - \ket{11})/\sqrt{2}, \\
    \ket{\sigma_{11}} = (\ket{01} - \ket{10})/\sqrt{2}.
\end{align}
These Bell states form an orthonormal basis for two-qubit systems, known as the Bell basis. For ease of notation, we will write $\ket{\sigma_{r^z r^x}}= \ket{r^z,r^x}$. We denote the tensor product of Bell states as $\ket{\mathbf{r}}=\otimes_i\ket{r^z_i,r^x_i}$, where $\mathbf{r}=\{\mathbf{r}^z,\mathbf{r}^x\}$. 

\begin{figure}[ht!]
\centering
\includegraphics[width=.6\linewidth]{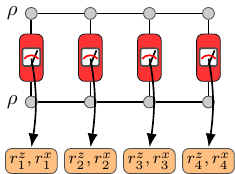}
\caption{Bell sampling from two copies of the state $\rho$. The measurement outcomes are $r_i^z$ and $r_i^x$ which label the Bell states in Eq.~\eqref{eq:bell_states}.}
\label{fig:bell_sampling}
\end{figure}

For two states $\rho_A, \rho_B$, measurements in the Bell basis are performed by preparing the tensor product state $\rho_A \otimes \rho_B$, and then applying the Bell transformation 
\begin{equation}
    U_\mathrm{Bell} = \bigotimes_{i=1}^N (\mathrm{H}\otimes {I}_2) \mathrm{CNOT} 
\end{equation}
 with the Hadamard gate $\mathrm{H}=\frac{1}{\sqrt{2}}(X+Z)$ and the CNOT gate $\text{CNOT}=\sum_{a_1,a_2}\ket{a_1,a_2\otimes a_1}\bra{a_1,a_2}$. We can then measure in the computational basis, yielding a measurement outcome $\mathbf{r}=\{\mathbf{r}^z, \mathbf{r}^x\}$ with $\mathbf{r}^z, \mathbf{r}^x\in \{0,1 \}^{N}$ with probability
\begin{equation}
    P(\mathbf{r}) = \bra{\mathbf{r}} \rho \otimes \rho  \ket{\mathbf{r}} = \frac{1}{2^N} \Tr[\rho_A \sigma_\mathbf{r} \rho_B^* \sigma_\mathbf{r} ],
\end{equation}
where $\sigma_\mathbf{r} = \otimes_i \sigma_{r^z_i r^x_i}$ is the $N$-qubit Pauli operator corresponding to the outcome $\mathbf{r}$. It is schematically illustrated in Fig.~\ref{fig:bell_sampling}. 

For a pure state $\ket{\psi}$ of $N$ qubits, performing Bell measurement on the two copies $\ket{\psi} \otimes \ket{\psi}$ yields a measurement outcome $\mathbf{r}^z, \mathbf{r}^x \in \{0,1 \}^{N}$ with probability
\begin{equation} \label{eq:pauli_dist}
    P(\mathbf{r}) = \frac{1}{2^N} | \bra{\psi} \sigma_\mathbf{r} \ket{\psi^*}|^2,
\end{equation}
where $\ket{\psi^*}$ is the complex conjugate of $\ket{\psi}$. If $\ket{\psi}$ has real coefficients, or if one measures from $\ket{\psi} \otimes \ket{\psi^*}$, the probability above becomes $\Xi(\mathbf{r}) = | \bra{\psi} \sigma_\mathbf{r} \ket{\psi}|^2 / 2^N$, which is commonly known as the characteristic function~\cite{gross2021schur}, a probability distribution over the set of $4^N$ (unsigned) Pauli strings.

\section{Update scheme for the transverse field Ising model} 
\label{sec:update_tfim}

We recall the operators
\begin{align} \label{eq:operators}
    &\HH_{0,i} =  I_i \otimes I_i , \\
    &\HH_{1,i} =  \XX_i, \\
    &\HH_{2,\langle i,j \rangle} =  I_i I_j \otimes I_i I_j , \\
    &\HH_{3,\langle i,j \rangle} =  \ZZ_{\langle i,j \rangle},
\end{align}
where we call $\HH_0$ and $\HH_1$ as site operators, while $\HH_2$ and $\HH_3$ are bond operators. Moreover, $\HH_0$ and $\HH_2$ are diagonal operators (in the Bell basis), while $\HH_1$ and $\HH_3$ are off-diagonal. The two-copy Hamiltonian for the TFIM can be written as 
\begin{equation}
    \HH = -2h\sum_{t=0,1} \sum_i \HH_{t,i} - 2\sum_{t=2,3} \sum_{\langle i,j \rangle} \HH_{t,{\langle i,j \rangle}},
\end{equation}
up to a constant energy shift. 

\begin{figure}[ht!]
    \centering
    \begin{overpic}[width=0.65\linewidth]{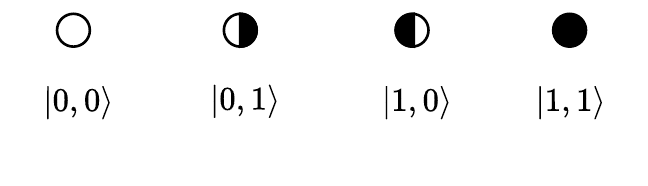}
    \put (-15,23) {{\textbf{(a)}}}
    \end{overpic}
    \begin{overpic}[width=0.8\linewidth]{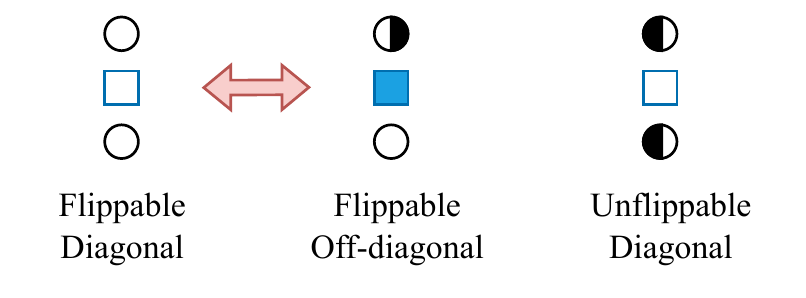}
    \put (-8,25) {{\textbf{(b)}}}
    \end{overpic}
    \begin{overpic}[width=0.8\linewidth]{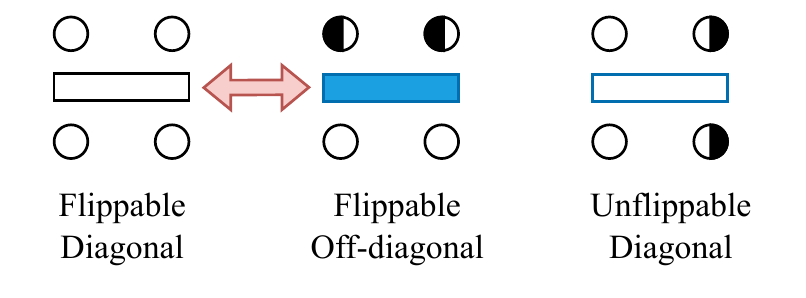}
    \put (-8,25) {{\textbf{(c)}}}
    \end{overpic}
    \caption{
    (a) The state $\ket{r^z, r^x}$ is represented by a circle, with the left half encoding $r^z$ and the right half $r^x$; an empty half denotes 0, and a filled half denotes 1.
    (b) A site operator is denoted by a square (empty for diagonal, filled for off-diagonal), where the matrix element $\langle r_i'^z ,r_i'^x|\mathcal{H}_{t,i}|r_i^z, r_i^x\rangle$ is represented as shown. From left to right, the three examples correspond to $\langle 0,0|\mathcal{H}_{0,i}|0,0\rangle$, $\langle 0,0|\mathcal{H}_{1,i}|0,1\rangle$, and $\langle 1,0|\mathcal{H}_{0,i}|1,0\rangle$, respectively.
    (c) A bond operator is denoted by a rectangle (empty for diagonal, filled for off-diagonal), where the matrix element $\langle r_i'^z, r_i'^x| \langle r_j'^z, r_j'^x|\mathcal{H}_{t,\langle ij\rangle}|r_i^z, r_i^x\rangle | r_j^z, r_j^x\rangle$ is represented as shown. From left to right, the three examples correspond to $\langle 0,0| \langle 0,0|\mathcal{H}_{0,\langle ij\rangle}|0,0\rangle |0,0\rangle$, $\langle 0,0|\langle0,0|\mathcal{H}_{1,\langle ij\rangle}|1,0\rangle|1,0\rangle$, and $\langle 0,0|\langle0,1|\mathcal{H}_{0,\langle ij\rangle}|0,0\rangle|0,1\rangle$, respectively.
    } 
    \label{fig:graph_rep}
\end{figure}

\begin{figure}[ht!]
    \centering
    \begin{overpic}[width=0.48\linewidth]{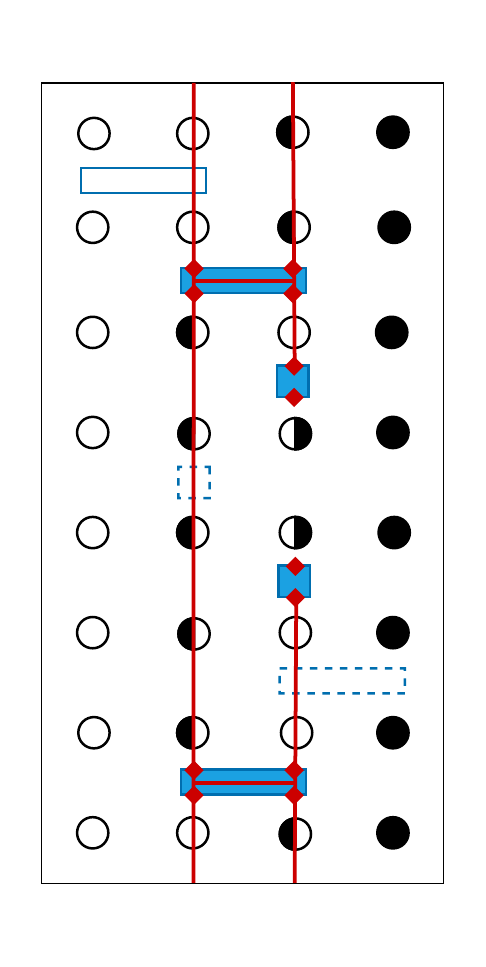}
    \put (22.5,-1) {{\textbf{(a)}}}
    \end{overpic}
    \begin{overpic}[width=0.48\linewidth]{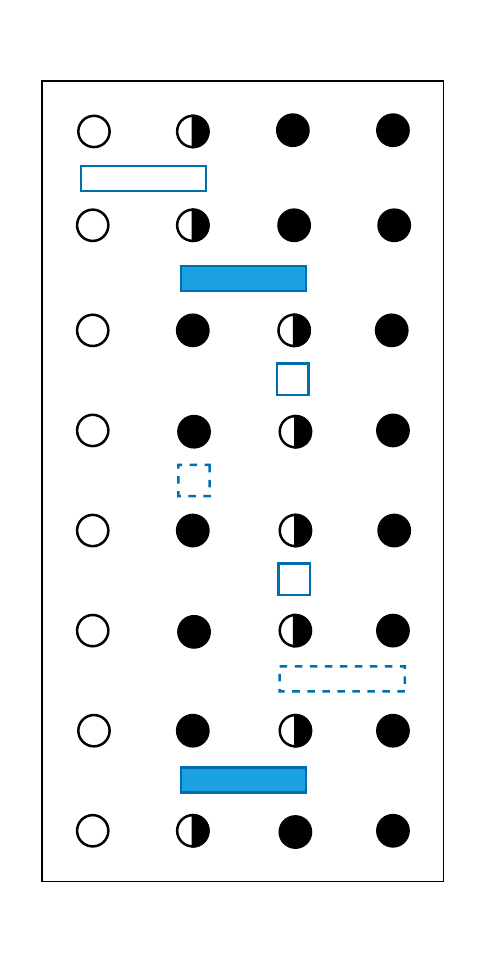}
    \put (22.5,-1) {{\textbf{(b)}}}
    \end{overpic}
    \caption{Configurations from a Bell-SSE simulation of a four-site 1D TFIM. 
    Lattice sites are arranged horizontally. Along the imaginary-time (vertical) direction, states are propagated by operators represented by rectangles.
    Dashed rectangles represent unflippable operators excluded from both cluster and bond cluster updates. Operator legs are marked by red diamonds, and operators within the same cluster or bond cluster are connected by red lines. 
    (a) Configuration before flipping the cluster;
    (b) Configuration after flipping the cluster.
    } 
    \label{fig:sse_configs_cluster}
\end{figure}

\begin{figure}[ht!]
    \centering
    \begin{overpic}[width=0.48\linewidth]{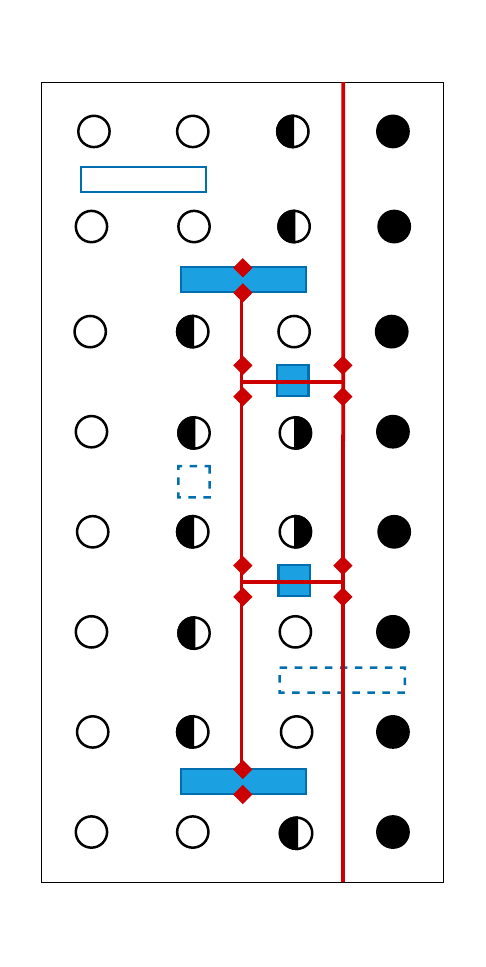}
    \put (22.5,-1) {{\textbf{(a)}}}
    \end{overpic}
    \begin{overpic}[width=0.48\linewidth]{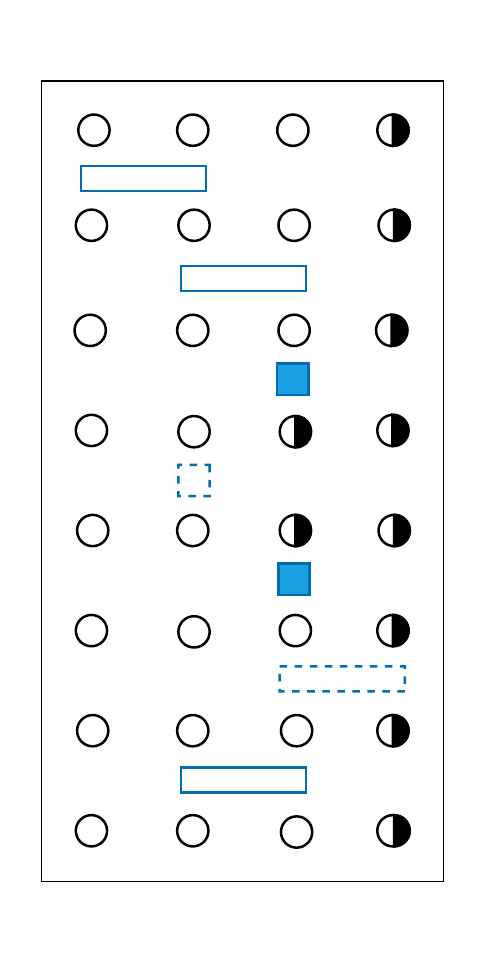}
    \put (22.5,-1) {{\textbf{(b)}}}
    \end{overpic}
    \caption{Configurations from a Bell-SSE simulation of a four-site 1D TFIM. 
    Lattice sites are arranged horizontally. Along the imaginary-time (vertical) direction, states are propagated by operators represented by rectangles.
    Dashed rectangles represent unflippable operators excluded from both cluster and bond cluster updates. Operator legs are marked by red diamonds, and operators within the same cluster or bond cluster are connected by red lines. 
    (a) Configuration before flipping the bond cluster;
    (b) Configuration after flipping the bond cluster.
    } 
    \label{fig:sse_configs_bond_cluster}
\end{figure}

In Fig.~\ref{fig:graph_rep}, we present a graphical representation of the states and operators, analogous to that used in standard SSE formulations on the Pauli-$Z$ basis~\cite{sandvik2003sse_ising,sandvik2019sse,Melko2013sse}, which serves as the foundation for the updates that follow.

To perform the cluster update, we introduce the linked vertex representation~\cite{sandvik1999sse_loop}, in which each operator is represented by a vertex with multiple legs. As illustrated in Fig.~\ref{fig:sse_configs_cluster}(a), a site operator has two legs, while a bond operator has four legs, all of which are connected to site variables.

Moreover, in order to preserve the constraints in Eq.~\eqref{eq:action_x} and \eqref{eq:action_z}, we divide diagonal operators into two types: flippable and unflippable, as shown in Fig.~\ref{fig:graph_rep}(b). A diagonal operator is flippable if it can be flipped into an off-diagonal operator without breaking the constraints, and it is unflippable otherwise. For example, if $\HH_{0,i}$ acts on the Bell state $\ket{1,0}$, flipping the operator would result in vanishing weight due to Eq.~\eqref{eq:action_x}; thus, the operator is unflippable (see Fig.~\ref{fig:graph_rep}(b)). 

The construction of clusters thus follows the standard procedure~\cite{sandvik2003sse_ising}, with only off-diagonal and flippable diagonal site operators being considered, while off-diagonal bond operators induce branching of the cluster. Therefore, compared to the standard cluster update, off-diagonal bond operators play the role of bond operators, flippable diagonal and off-diagonal site operators play the role of site operators, and all other types of operators, i.e. diagonal bond operators and unflippable site operators, are ignored. Fig.~\ref{fig:sse_configs_cluster} shows an example of a cluster. This update flips the $r^x$ part of the Bell states and the site operators that are included in the cluster. 

\begin{algorithm}[h!]
\caption{Cluster Update in Bell-SSE}
\label{alg:cluster_update}
\KwIn{An SSE configuration with operators $\{\mathcal{H}_{t,b}\}$ and Bell states $\mathbf{r}$ at the initial imaginary-time.}
\KwOut{An updated configuration with $\{\mathcal{H}'_{t,b}\}$ and $\mathbf{r}'$.}

Mark all operator legs $\{v\}$ related to site variables as unvisited\;
Identify all flippable site operators and off-diagonal bond operators as \emph{target operators}\;

\For{leg $v = 0, 1, 2, \ldots$}{
    \If{$v$ is unvisited}{
        \If{$v$ belongs to a target operator}{
            Start a new cluster from $v$\;
            Grow the cluster by recursively adding all other connected, unvisited legs that belong to some target operators according to the rules illustrated in Fig.~\ref{fig:sse_configs_cluster}\;
            \If{a random number $\in[0,1]$ $\le 1/2$}{
                Flip the cluster, changing all $\mathcal{H}_{0,i}$ to $\mathcal{H}_{1,i}$ in the cluster, and vice versa\;
                Mark all legs in the cluster as visited and flipped\;
            }
            \Else{
                Mark all legs in the cluster as visited but unflipped\;
            }
        }
    }
}

\For{site $i = 0, 1, 2, \ldots$}{
    \If{no operator acts on $i$}{
        \If{a random number $\in[0,1]$ $\le 1/2$}{
            $r_i^x \gets r_i^x \oplus 1$\;
        }
    }
    \Else{
        \If{at least one target operator acts on $i$}{
            Find the nearest leg $v$ (in imaginary time, starting from the initial slice) associated with $i$ that belongs to a target operator\;
            \If{$v$ is marked as flipped}{
                $r_i^x \gets r_i^x \oplus 1$\;
            }
        }
    }
}
\end{algorithm}

The complete procedure for the cluster update of the 1D TFIM is presented in Algorithm~\ref{alg:cluster_update}. In practice, it suffices to record the Bell state $\mathbf{r}$ of all sites at the initial imaginary-time slice, as all other slices are uniquely determined once the operators $\{\mathcal{H}_{t,b}\}$, which propagate the states, are specified, where $b$ denotes a site variable or a bond variable.

To update the off-diagonal bond operators, the standard approach is by performing loop update~\cite{sandvik1999sse_loop}. However, if an off-diagonal site operator is encountered during a loop construction, the corresponding loop cannot be flipped as it would violate the constraint. Thus, the flipping of the loop has to be rejected, resulting in a highly inefficient update scheme. In practice, most of the loops cannot be flipped, thus breaking the ergodicity. To solve this issue, we introduce a new type of update that we call bond cluster update, since it can be seen as a cluster update that acts on the bond variables instead of the site variables. 

We start by introducing a dummy variable $\mu$ at each bond and impose the constraint $\mu^x_{\langle i-1,i\rangle} X_i \mu^x_{\langle i,i+1\rangle }=1$ at each site. The Hamiltonian becomes a $\Z_2$ gauge theory in 1D in the form $H=-\sum_i Z_i \mu^z_{\langle i,i+1\rangle} Z_{i+1}-h\sum_i \mu^x_{\langle i-1,i\rangle}\mu^x_{\langle i,i+1\rangle}$. One can see that the role of site operators and bond operators are exchanged for the bond variables $\mu$.

In the bond cluster update, the legs of a vertex are connected to the bond variables. A site operator $\XX_i$ acts on the bond variables at $\langle i-1,i\rangle$ and $\langle i,i+1\rangle$, while a bond operator $\ZZ_{\langle i,i+1\rangle}$ acts on the bond variable at $\langle i,i+1\rangle$. Therefore, a site operator has four legs, while a bond operator has two legs for the 1D TFIM, as illustrated in Fig.~\ref{fig:sse_configs_bond_cluster}. The update then proceeds as in the cluster update, with only off-diagonal and flippable diagonal bond operators being considered, while off-diagonal site operators induce branching of the bond cluster (see Fig.~\ref{fig:graph_rep}(c) for examples of flippable and unflippable bond operators). Note that in practice, the bond variables do not actually need to be stored. The update simply flips the $r^z$ part of the Bell states on the sites that are connected by the bonds, as well as the bond operators that are included in the bond cluster. Note that the parity $\oplus_{i=1}^N r^z_i$ is preserved in this update. This is in fact consistent with the $\Z_2$ symmetry in the TFIM, which is generated by $\prod_i X_i$. In 1D, the cluster update and bond cluster update are related by the Kramers-Wannier duality, which exchanges $X$ and $ZZ$.  
Algorithm~\ref{alg:bond_cluster_update} presents the complete procedure for the bond-cluster update in the 1D TFIM. The final step, which simultaneously updates the $r^x$ component of all site states, is included to enhance ergodicity without violating the constraints.

The bond cluster update is a new type of update that may be of independent interest, as it can also be applied in the standard SSE to simulate the TFIM in the $X$ basis, which has typically been simulated using the directed loop algorithm~\cite{syljusen2002directedloops,syljusen2003directed,zhao2021higherform}. Note that the Bell-QMC algorithm in the TFIM appears similar to simulating the $Z$ and $X$ basis at the same time, yet it provides an exponential gain in the range of accessible operators over conventional QMC. 

\begin{algorithm}[h!]
\caption{Bond Cluster Update in Bell-SSE}
\label{alg:bond_cluster_update}
\KwIn{An SSE configuration with operators $\{\mathcal{H}_{t,b}\}$ and Bell states $\mathbf{r}$ at the initial imaginary-time.}
\KwOut{An updated configuration with $\{\mathcal{H}'_{t,b}\}$ and $\mathbf{r}'$.}

Mark all operator legs $\{v\}$ related to bond variables as unvisited\;
Identify all flippable bond operators and off-diagonal site operators as \emph{target operators}\;

\For{leg $v = 0, 1, 2, \ldots$}{
    \If{$v$ is unvisited}{
        \If{$v$ belongs to a target operator}{
            Start a new bond cluster from $v$\;
            Grow the bond cluster by recursively adding all other connected, unvisited legs that belong to some target operators according to the rules illustrated in Fig.~\ref{fig:sse_configs_bond_cluster}\;
            \If{a random number $\in[0,1]$ $\le 1/2$}{
                Flip the bond cluster, changing all $\mathcal{H}_{2,i}$ to $\mathcal{H}_{3,i}$ in the bond cluster, and vice versa\;
                Mark all legs in the bond cluster as visited and flipped\;
            }
            \Else{
                Mark all legs in the bond cluster as visited but unflipped\;
            }
        }
    }
}

\For{bond $\langle ij\rangle = \langle 01\rangle, \langle 12\rangle, \langle 23\rangle, \ldots$}{
    \If{no operator acts on $\langle ij\rangle$}{
        \If{a random number $\in[0,1]$ $\le 1/2$}{
            $r_i^z \gets r_i^z \oplus 1$\;
            $r_j^z \gets r_j^z \oplus 1$\;
        }
    }
    \Else{
        \If{at least one target operator acts on $\langle ij\rangle$}{
            Find the nearest leg $v$ (in imaginary time, starting from the initial slice) associated with $\langle ij\rangle$ that belongs to a target operator\;
            \If{$v$ is marked as flipped}{
                $r_i^z \gets r_i^z \oplus 1$\;
                $r_j^z \gets r_j^z \oplus 1$\;
            }
        }
    }
}

\If{a random number $\in[0,1]$ $\le 1/2$}{
    \For{site $i=0,1,2,\ldots$}{
        $r_i^x \gets r_i^x \oplus 1$\;
    }
}

\end{algorithm}

While we have focused on 1D TFIM, we note that the update scheme is general for nonfrustrated Ising interactions. For example, with next-nearest-neighbor interaction in 1D, the operator has four legs located on the two bonds between the two sites. However, we note that these updates are not ergodic in 1D with periodic boundary condition (PBC), as configurations with bond operators winding along the periodic direction cannot be generated. Nevertheless, we find that the results still recover the exact values even for simulations with PBC. If desired, the ergodicity can be restored by considering a ring update as discussed in~\cite{sandvik1997finitesize}. In 2D and higher, an additional type of update which flips bond operators around a closed loop is also required.

\section{Update scheme for $\Z_2$ lattice gauge theory} \label{sec:update_z2_lgt}
The update scheme for $\Z_2$ LGT is analogous to that of the TFIM described in the previous section, with appropriate modification. We decompose the Hamiltonian of $\Z_2$ LGT as
\begin{equation} \label{eq:z2_lgt_2copy}
    \HH = - 2\sum_{\square} \XX_{\square}  - 2h\sum_i \ZZ_i  ,
\end{equation}
where 
\begin{align}
    & \XX_{\square} = \frac{1}{2}(\prod_{i\in\square} X_i \otimes \prod_{i\in\square} I_i + \prod_{i\in\square} I_i \otimes \prod_{i\in\square} X_i) \\ 
    & \ZZ_i = \frac{1}{2}(Z_i \otimes I_i + I_i  \otimes Z_i ).
\end{align}
The action of these terms on a Bell state is given by
\begin{equation} \label{eq:action_x_z2_lgt}
    \XX_\square \bigotimes_{i\in\square}\ket{r^z_i,r^x_i} = \delta_{\oplus_{i\in\square}r^z_i,0}\bigotimes_{i\in\square}\ket{r^z_i,r^x_i\oplus1}, 
\end{equation}
and 
\begin{equation} \label{eq:action_z_z2_lgt}
    \ZZ_i \ket{r^z_i,r^x_i}  = \delta_{r^x_i  ,0} \ket{r^z_i  \oplus 1,r^x_i} .
\end{equation}

We then define the operators
\begin{align} \label{eq:operators_z2_lgt}
    &\HH_{0,i} =  I_i \otimes I_i , \\
    &\HH_{1,i} =  \ZZ_i, \\
    &\HH_{2,\square} =  \prod_{i\in\square} I_i \otimes \prod_{i\in\square} I_i , \\
    &\HH_{3,\square} =  \XX_{\square}. 
\end{align}
$\HH_0$ and $\HH_1$ are site operators, while $\HH_2$ and $\HH_3$ are plaquette operators. Moreover, $\HH_0$ and $\HH_2$ are diagonal operators, while $\HH_1$ and $\HH_3$ are off-diagonal. The $\Z_2$ LGT in Eq.~\eqref{eq:z2_lgt_2copy} can thus be written as 
\begin{equation}
    \HH = -2h\sum_{t=0,1} \sum_i \HH_{t,i} - 2\sum_{t=2,3} \sum_{\square} \HH_{t,\square}
\end{equation}
up to a constant energy shift.

We again use three types of updates: (1) diagonal update, (2) cluster update, and (3) plaquette cluster update. The cluster update is used to change between $\HH_{0,i}$ and $\HH_{1,i}$, while the plaquette cluster update changes between $\HH_{2,b}$ and $\HH_{3,b}$.

In the cluster update, we introduce the legs connected to the sites where the operators act. Thus, a site operator has two legs, and a plaquette operator has eight legs. Similarly as in the TFIM, a cluster branches when encountering a plaquette operator and terminates at a site operator. However, a naive approach of branching the cluster on all eight legs of the plaquette operator leads to the proliferation problem~\cite{melko2005stochastic}, where clusters tend to encompass a significant portion of the lattice, severely hindering the effectiveness of the update. To solve this, in each MC step and for each plaquette, we randomly divide the four links into two groups. Then, the original eight-leg vertex is decomposed into two disconnected sub-vertices, each of which includes four legs.  This effectively decomposes the cluster that involves eight-leg vertices into sub-clusters during the updates, which enhances the ergodicity.

In the plaquette cluster update, the legs are connected to plaquettes, where a site operator has four legs and a plaquette operator has two legs. Thus, a plaquette cluster branches when encountering a site operator, and it terminates at a plaquette operator.

\begin{figure}
\centering
\includegraphics[width=.5\linewidth]{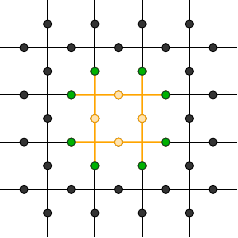}
\caption{The spins in the square partition divided into two groups: boundary spins (green) in $\partial A$ and bulk spins (yellow) in $A / \partial A$.}
\label{fig:z2_lgt_ent2}
\end{figure}

\section{Exploiting symmetry} \label{sec:symmetry}
Exploiting the symmetries of the model is crucial for improving the statistics and reducing the number of measurements required.

First, the lattice symmetries, such as translation and rotation, can be exploited in a simple way by averaging measurements of all observables that are equivalent by symmetry.

Second, if the Hamiltonian has some conserved quantum numbers, such as $\Z_2$ or $U(1)$, then the ground state typically belongs to a specific symmetry sector. In the case of the TFIM, the ground state belongs to the sector $P=1$, where $P=\prod_i X_i$. This has two effects: (1) We only need to consider Bell states satisfying the condition $\oplus_i r^z_i = 0$, and (2) Bell states that differ by $\oplus_i r^x_i$ are equivalent by symmetry.

Similar considerations apply for the $\Z_2$ LGT, where the symmetry arises from the Gauss' law. Notably, the estimator for the entanglement averaged over all the gauge transformations can be determined analytically, yielding
\begin{equation} 
    S_2(A) = - \ln \left\langle \prod_{i\in \partial A} \delta_{r_i^x,0} \prod_{i\in A / \partial A} (-1)^{r^x_i r^z_i} \right\rangle_{\mathbf{r}},
\end{equation}
 where $\partial A$ is the set of spins at the boundary of the partition (see Fig.~\ref{fig:z2_lgt_ent2}).

\section{Simulations of the extended ensembles}\label{sec:sim_extended_ensmble}
For the extended ensemble 
\begin{equation}
    Q(\lambda) = \sum_{B \subseteq A} \lambda^{N_B} (1-\lambda)^{N_A - N_B} Q_B,
\end{equation}
the configuration space is divided into subspaces labeled by each subset $B\subseteq A$. To ensure proper sampling across these subspaces, during each Monte Carlo step, we traverse the sites in region $A$ sequentially and attempt the following updates:
\begin{enumerate}[label=(\alph*)]
    \item If site $j$ is currently in $B$ and is in the state $\ket{0,0}$, we propose removing it from $B$ with the probability
    \begin{equation}
        p_{\mathrm{remove}}=\min\{1,\frac{1-\lambda}{\lambda}\}.
    \end{equation}
    A site in $B$ is allowed to be updated during both cluster and bond cluster updates at the zero imaginary time. If the site is removed, it is fixed in the state $\ket{0,0}$ in the subsequent step.
    \item If site $j$ is not in $B$, and hence must be in the state $\ket{0,0}$, we propose adding it to $B$ with probability 
    \begin{equation}
        p_{\mathrm{add}}=\min\{1, \frac{\lambda}{1-\lambda}\}.
    \end{equation}
\end{enumerate}

In the simulation of 
\begin{equation}
Q(\lambda)=\sum_{P\in\mathcal{P}_A}\lambda^{\mathrm{wt}(P)}|\Tr(e^{-\beta H} P)|^2,
\end{equation}
which corresponds to an analytical average over subsets $B \subseteq A$, the region outside $A$ is kept fixed at $\ket{00}$ throughout. Consequently, during off-diagonal updates, clusters and bond clusters that attempt to flip the site at zero imaginary time needs to be modified. The acceptance probability is given by
\begin{equation}
    p=\min\{1, \lambda^{\mathrm{wt}(P')-\mathrm{wt}(P)}\},
\end{equation}
where $P$ and $P'$ are the Pauli strings before and after the proposed (bond) cluster flip, respectively.

\section{Variance comparison of the estimators} \label{sec:var_compare}
\begin{figure}[ht!]
\centering
\begin{overpic}[width=0.48\linewidth]{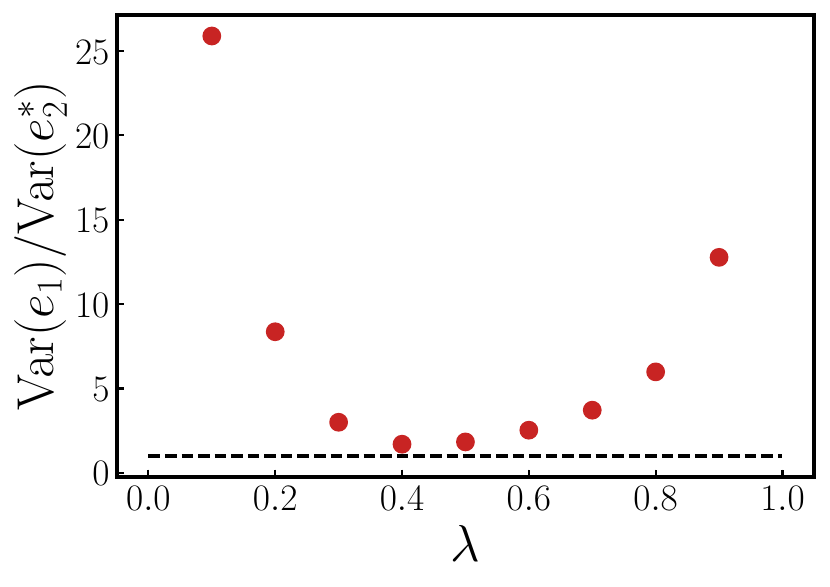}
\put (-3,75) {{\textbf{(a)}}}
\end{overpic}
\begin{overpic}[width=0.48\linewidth]{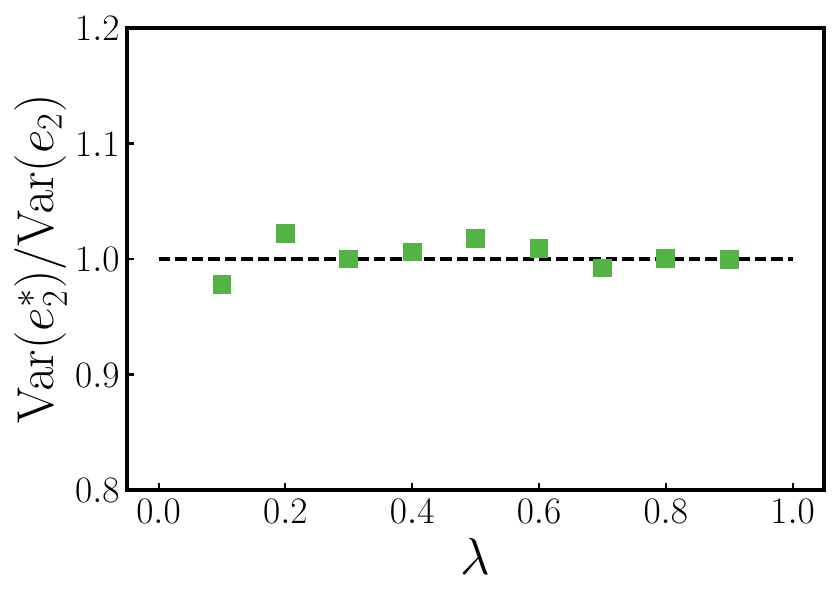}
\put (-3,75) {{\textbf{(b)}}}
\end{overpic}
\caption{An example of 2D $\mathbb{Z}_2$ LGT with $L=10$ and $h=0.3$, for the variance comparison of evaluating (a) $\langle e_1\rangle_{\lambda}$ and $\langle e_2^*\rangle_{\lambda}$, (b) $\langle e_2^*\rangle_{\lambda}$ and $\langle e_2\rangle_{\lambda}$. The dashed horizontal line denotes the value of one.}
\label{fig:raw_variance_comparison}
\end{figure}

Since Eq.~\eqref{eq:extended_ensemble} and Eq.~\eqref{eq:new_extended_ensemble} are equivalent, we can also estimate $\langle e_2\rangle_{\lambda}$ by simulating Eq.~\eqref{eq:extended_ensemble}, and we denote this alternative estimator by $\langle e_2^{*}\rangle_{\lambda}$.  Fig.~\ref{fig:raw_variance_comparison} shows an example of comparing the raw variances of $e_1$, $e_2$, and $e_2^*$, taking into account the distinct autocorrelations associated with the simulations of Eq.~\eqref{eq:extended_ensemble} and Eq.~\eqref{eq:new_extended_ensemble}. The significantly reduced variances of both $e_2$ and $e_2^*$ compared to that of $e_1$ demonstrate the effectiveness of the proposed formulation in Eq.~\eqref{eq:new_extended_ensemble} and estimators $\langle e_2^*\rangle_{\lambda}$ and $\langle e_2\rangle_{\lambda}$. 
The variance reduction achieved by averaging over region $B$ is particularly crucial when applying the non-equilibrium Jarzynski equation. In each quench process, only a single sample is used to compute the non-equilibrium work for every value of $\lambda$~\cite{emidio2020entanglement}. In such scenarios, employing Eq.~\eqref{eq:new_extended_ensemble} should offer a marked improvement in data quality over Eq.~\eqref{eq:extended_ensemble}.

\begin{figure}
\centering
\includegraphics[width=.52\linewidth]{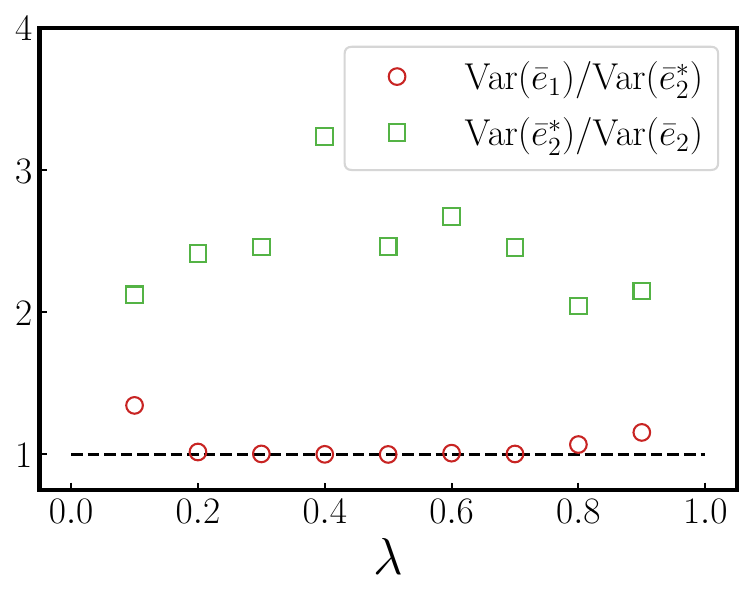}
\caption{
An example of 2D $\mathbb{Z}_2$ LGT with $L=10$ and $h=0.3$, for the variance comparison of evaluating the binned averages $\langle \bar e_1\rangle_{\lambda}$, $\langle \bar e_2^*\rangle_{\lambda}$, and $\langle \bar e_2\rangle_{\lambda}$. The dashed horizontal line denotes the value of one.
}
\label{fig:binned_variance_comparison}
\end{figure}

Because the raw variances are large, we then consider the binned averages—$ \bar{e}_1 $, $\bar{e}_2^* $, and $\bar{e}_2 $—to further compare the estimators, especially $e_2^* $ and $ e_2$, whose differences are not sufficiently clear in Fig.~\ref{fig:raw_variance_comparison}b. Binning is a standard technique in equilibrium Monte Carlo sampling, and we adopt a block size of 5000 to generate effectively uncorrelated averages. As shown in Fig.~\ref{fig:binned_variance_comparison}, using $ \bar{e}_2^* $ also yields smaller variances than using $ \bar{e}_1 $, and the tendency is consistent with that in Fig.~\ref{fig:raw_variance_comparison}a. The advantage of $\bar{e}_2 $ over $\bar{e}_2^*$ is also more evident after binning, highlighting the strength of Eq.~\eqref{eq:new_extended_ensemble}. We found similar results on smaller lattices.

\section{Benchmark with exact diagonalization}
\label{sec:benchmark}
\begin{figure} \label{fig:benchmark_ising}
\centering
\begin{overpic}[width=0.48\linewidth]{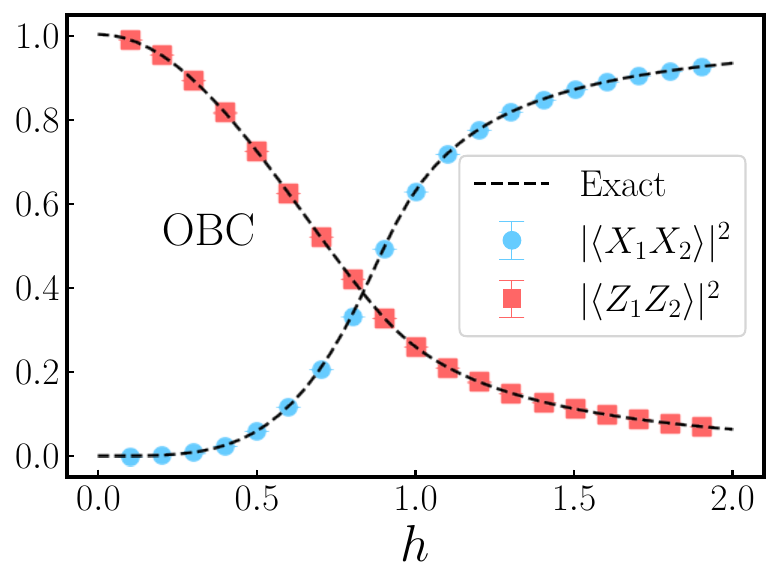}
\put (-3,75) {{\textbf{(a)}}}
\end{overpic}
\begin{overpic}[width=0.48\linewidth]{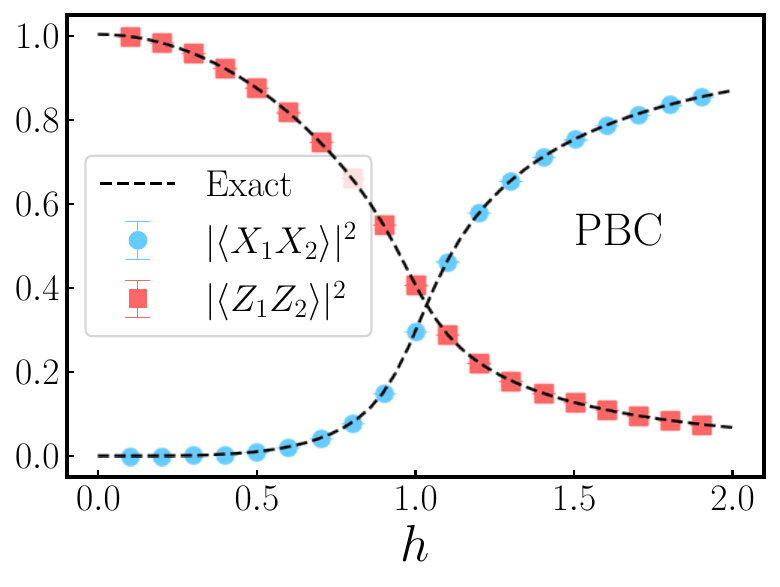}
\put (-3,75) {{\textbf{(b)}}}
\end{overpic}
\caption{Benchmark of $\vert \langle X_1 X_2 \rangle\vert^2$ and $\vert\langle Z_1 Z_2\rangle\vert^2$ with exact diagonalization in the 1D TFIM with (a) open and (b) periodic boundary condition. The system size is $L=20$.}
\end{figure}

\begin{figure}
\centering
\includegraphics[width=.52\linewidth]{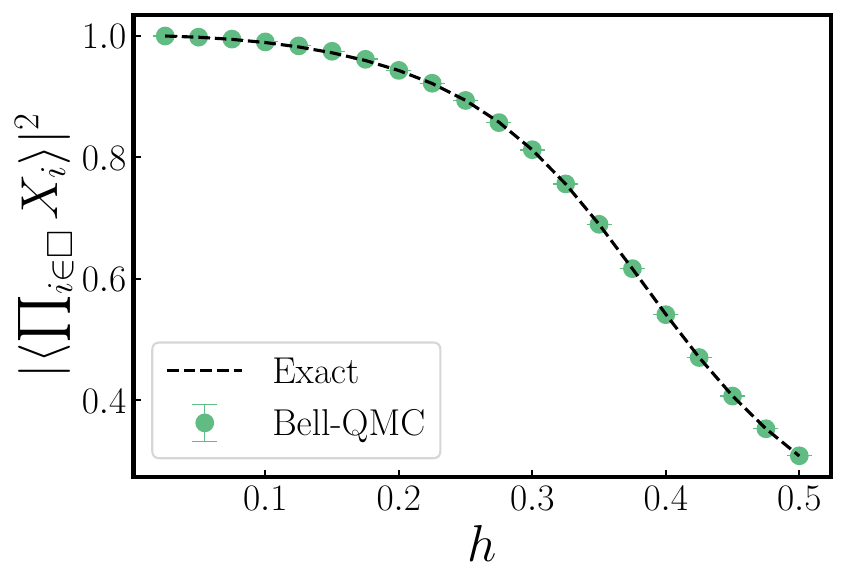}
\caption{Benchmark of the Wilson loop $\vert \langle \prod_{i\square} X_i \rangle\vert^2$ with exact diagonalization in the 2D $\Z_2$ LGT on a $3\times 3$ lattice.}
\label{fig:benchmark_z2}
\end{figure}

In this section, we present benchmark results comparing our Bell-QMC method against exact diagonalization for both the 1D TFIM and $\Z_2$ LGT. In Fig.~\ref{fig:benchmark_ising}, we show the results of $\vert \langle X_1 X_2 \rangle\vert^2$ and $\vert\langle Z_1  Z_2\rangle\vert^2$ calculated using Bell-QMC, where subscripts indicate the first and second lattice sites. The excellent agreement with exact diagonalization confirms the correctness of our algorithm. Then, Fig.~\ref{fig:benchmark_z2} shows the Wilson loop operator in the $\Z_2$ LGT, again demonstrating a perfect agreement with exact diagonalization results.

\section{Additional numerical data}
\begin{figure}
\centering
\begin{overpic}[width=0.48\linewidth]{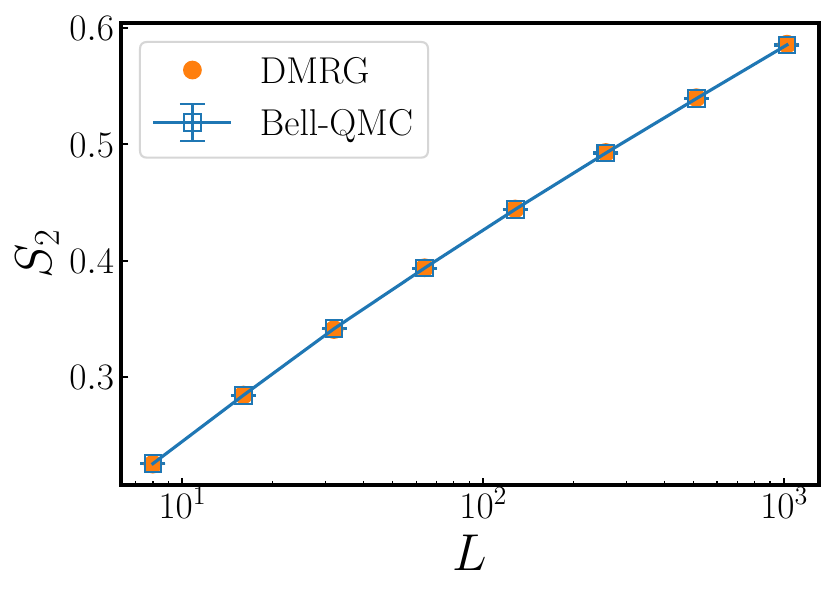}
\put (-3,70) {{\textbf{(a)}}}
\end{overpic}
\begin{overpic}[width=0.48\linewidth]{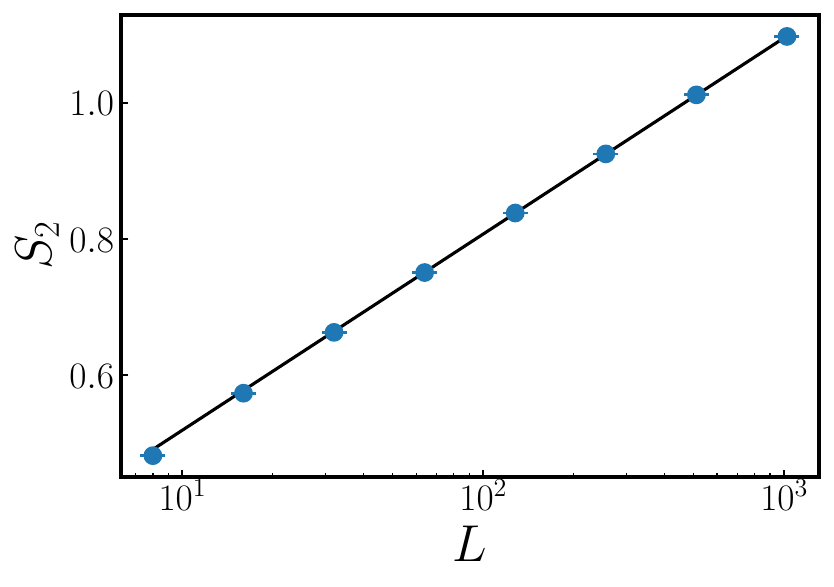}
\put (-3,70) {{\textbf{(b)}}}
\end{overpic}
\caption{ R\'enyi-2 entanglement $S_2$ in the ground state of the TFIM at the critical point $h=1$ in the half bipartition. In (a), we benchmark $S_2$ with DMRG for various system sizes up to $L=1024$ with open boundary condition and $\beta=3L$. In (b), we show $S_2$ in systems with periodic boundary condition with $\beta=4L$. The solid line denotes a linear fit from $L \in \{64, 128, 256,512,1024\}$, from which we extract the central charge $c \approx 0.4999(3)$ }
\label{fig:ising_half_chain}
\end{figure}

Here, we present additional numerical data on the TFIM.
We consider open boundary condition to benchmark with DMRG simulations. In DMRG, the entanglement is directly available for any left-right cut, and we choose the half bipartition of the chain. We focus on the critical point $h_c=1$. The data we obtain using Bell-QMC are in excellent agreement with the DMRG results, as shown in Fig.~\ref{fig:ising_half_chain}a. We further simulate systems with periodic boundary condition, showing clear logarithmic scaling. By fitting with the CFT scaling, $S_2 = \frac{c}{4} \ln \ell + \gamma$, we extract the central charge $c \approx 0.4999(3)$, as shown in Fig.~\ref{fig:ising_half_chain}b.

\end{document}